\documentclass[twocolumn]{autart}    % Enable this line and disable the 
\usepackage{cite}
\usepackage{amsfonts}
\usepackage{amsmath}
\usepackage{color}
\usepackage{textcomp}
\usepackage{graphics, graphicx}
\usepackage{enumerate}
\usepackage{array}
\usepackage{bm}
\usepackage{url} % not crucial - just used below for the URL 
\usepackage{verbatim}
\usepackage{multirow}
\usepackage{mathrsfs}
\usepackage{algorithm}
\usepackage{amssymb}
\usepackage{stfloats}
\usepackage{algpseudocode}
\usepackage{booktabs}

\newcommand*{\Scale}[2][4]{\scalebox{#1}{$#2$}}%

\newtheorem{mytheo}{Theorem}

\newtheorem{myrema}{Remark}
\newtheorem{mylemma}{Lemma}

\newtheorem{myassump}{Assumption}

\makeatletter
\let\NAT@parse\undefined
\makeatother

\begin{document}

\begin{frontmatter}

\title{Probabilistic reachable sets of stochastic nonlinear systems with contextual uncertainties\thanksref{footnoteinfo}} % Title, preferably not more 

\thanks[footnoteinfo]{This paper was not presented at any IFAC 
meeting. Corresponding author Y.~Wang.} %Tel. +XXXIX-VI-mmmxxi. 
%Fax +XXXIX-VI-mmmxxv.}

\author[Osaka]{Xun Shen}\ead{shenxun@eei.eng.osaka-u.ac.jp},    % Add the 
\author[Melbourne]{Ye Wang}\ead{ye.wang@unimelb.edu.au},               % e-mail address 
\author[Osaka]{Kazumune Hashimoto}\ead{hashimoto@eei.eng.osaka-u.ac.jp},
\author[Dalian]{Yuhu Wu}\ead{wuyuhu@dlut.edu.cn},  % (ead) as shown
\author[NTNU]{Sebastien Gros}\ead{sebastien.gros@ntnu.no}  % (ead) as shown

\address[Osaka]{Graduate School of Engineering, Osaka University, Osaka, 565-0871, Japan}  % Please supply                                   
\address[Melbourne]{School of Mathematics and Statistics, The University of Melbourne, VIC, 3010, Australia}             % full addresses
\address[Dalian]{Key Laboratory of Intelligent Control and Optimization for Industrial Equipment of Ministry of Education, School of Control Science and Engineering, Dalian University of Technology, Dalian,
116024, China}        % here.
\address[NTNU]{Center for Autonomous Marine Operations and Systems and Department of Engineering Cybernetics, Norwegian University of Science and Technology (NTNU), Trondheim, NO-7491 Norway}        % here.

\begin{keyword}                           % Five to ten keywords,  
Uncertainty quantification, probabilistic reachable sets, conditional probability, stochastic nonlinear systems, stochastic optimization.              % chosen from the IFAC 
\end{keyword}                             % keyword list or with the 
                                          % help of the Automatica 
                                          % keyword wizard

\begin{abstract}                          % Abstract of not more than 200 words.
Validating and controlling safety-critical systems in uncertain environments necessitates probabilistic reachable sets of future state evolutions. 
The existing methods of computing probabilistic reachable sets normally assume that stochastic uncertainties are independent of system states, inputs, and other environment variables. However, this assumption falls short in many real-world applications, where the probability distribution governing uncertainties depends on these variables, referred to as \emph{contextual uncertainties}. This paper addresses the challenge of computing probabilistic reachable sets of stochastic nonlinear states with contextual uncertainties by seeking minimum-volume polynomial sublevel sets with contextual chance constraints. 
The formulated problem cannot be solved by the existing sample-based approximation method since the existing methods do not consider conditional probability densities. 
To address this, we propose a consistent sample approximation of the original problem by leveraging conditional density estimation and resampling. The obtained approximate problem is a tractable optimization problem. Additionally, we prove the proposed sample-based approximation's almost uniform convergence, showing that it gives the optimal solution almost consistently with the original ones. Through a numerical example, we evaluate the effectiveness of the proposed method against existing approaches, highlighting its capability to significantly reduce the bias inherent in sample-based approximation without considering a conditional probability density.
\end{abstract}

\end{frontmatter}

%%%%%%%%%%%%%%%%%%%%%%%%%%%%%%%%%%%%%%%%%%%%%%%%%%%
%%%%%%%%%%%%%%%%%%%%%%%%%%%%%%%%%%%%%%%%%%%%%%%%%%%
\section{Introduction}
The future state evolution of a dynamical system under stochastic uncertainty may have infinitely possible realizations \cite{Haesaert}. 
Instead of accessing the safety of a specific scenario of the future state trajectory, it is necessary to clarify the safety of a set of future state trajectories, thereby ensuring safety from a probabilistic viewpoint. 
A probabilistic reachable set of future state trajectories specifies a confidence region, indicating that future states are located with a given probability level \cite{Shen2023_1, Shen2023_2, Mirko:2021}. 
Explicitly expressing probabilistic reachable sets is crucial for the design of robust controllers \cite{Karimi, Bloemers}, safety-critical validation \cite{Zanon, Akella:2022}, and anomaly detection \cite{A:Wan}. 
Computing probabilistic reachable sets can be approached as a chance-constrained optimization problem \cite{Dabbene:2015}. 
However, the robustness of constraints adds a layer of complexity to the chance-constrained optimization problem, which makes the search for solutions particularly difficult. 
This complexity underscores the challenges faced in seeking solutions \cite{Calafiore:2006, Margellos, CampiInterval, Shen2023_TNNLS}.

To tackle the challenging issue of intractable problems, approximation methods have been proposed to find approximate solutions to the original problem. 
Among these methods, the set-based approach has been proposed to compute probabilistic reachable sets in \cite{Au:Alamo}. 
For uncertain linear discrete-time systems, the linear matrix inequality technique has been applied to develop a method for computing the ellipsoidal boundaries of future state evolution in \cite{OE:Kishida}. 
In \cite{Dabbene:2015}, a method for computing the probabilistic bound of general nonlinear systems has been proposed based on the scenario approach in \cite{Calafiore:2006, Au:Alamo, Shen2023_1}. 
However, the method based on the standard scenario approach gives the solution that converges to the robust optimization solution instead of the random problem \cite{Margellos}, which often yields a fully robust confidence region. 
This region, although useful, does not precisely cater to the desired probability level of the confidence region.

The sampling-and-discarding approach, an enhanced version of the standard scenario approach, offers a more refined approximation for chance-constrained optimization problems \cite{Campi:2011, Campi:2018, Garatti:2022, Romao2022, Wang:Auto2024}. 
This method is distinguished by its ability to establish an exact relation between the risk level and the number of removal samples, which presents a systematic algorithm for sample elimination \cite{Luedtke, Shen2023_JOTA}. 
Although it represents a significant improvement in approximating chance constraints, the computational burden of this method increases dramatically due to the necessity of solving scenario programs repeatedly to identify the support samples. 
To mitigate the computations, \cite{Shen2023_2, Shen2023_TNNLS} defines the minimum-volume polynomial sublevel set proposed in \cite{Magnani} in a probabilistic version and proposes a sample-based continuous approximate problem that can be solved with low computational complexity and preserve a proper approximation to chance-constrained optimization. 
Smooth functions inspire the sample-based continuous approximation to emulate the indicator function \cite{Chen1995, Geletu:2017}, thereby transforming a mixed-integer program into a nonlinear continuous program. 

A significant limitation of the above existing methods is their assumption of independence between uncertainties and system states.  However, in various real-world applications, the probability distribution of uncertainties is influenced not only by the state \cite{Soloperto, Menner, Zhao_Pan} but also by other time-varying environment variables\footnote{Environment variables may include system inputs and/or other time-varying parameters in the system model.}.
In this work, we refer to such uncertainties as \textit{contextual uncertainties}. It is important to note that, as discussed in \cite{Kohler}, 
state- and input-dependent uncertainties are described as \emph{general disturbances}, which can be expressed as a function of states, inputs and previous disturbances. 
In contrast, contextual uncertainties cannot be expressed as a direct function of states and environment variables. Instead, the probability distribution governing these uncertainties is conditioned on these variables.
In these cases,  uncertainty samples are paired with state samples rather than distributed across different contexts or states. 
Directly using the uncertainty samples to construct sample-based approximations by the existing methods leads to a biased sample-based approximation of the chance constraints for each context since the samples are not extracted identically with the desired probability distribution of the corresponding context. 
Consequently, the solution to the approximate problem is not guaranteed to converge to the original one. 
Theoretical results that assess the quality of the solution to the approximate problem cannot be directly extended to the chance constraints with contextual uncertainties, as existing sample-based methods do not account for conditional probability densities. Distributionally chance-constrained optimization can improve performance in the worst case \cite{Arrigo}. 
However, the bias caused by the sample-based approximation cannot be eliminated. 
The information inside the uncertainty-state data pair should be utilized fully to improve the sample-based approximation of the chance constraints for the case with contextual uncertainties. 

This work addresses the challenge of computing probabilistic reachable sets for stochastic nonlinear systems with contextual uncertainties. The contributions of this paper are summarized as follows: Firstly, a novel formulation of the optimization problems to compute probabilistic reachable sets that account for contextual chance constraints is proposed. Then, approximate problems with contextual chance constraints using a conditional density estimation-based resampling strategy and establish theoretical results that quantify the solution quality. The proposed probabilistic reachable set computation algorithm employs conditional density estimation and a resampling strategy, drawing inspiration from machine learning and the concept of conditional density ratio estimation \cite{Biau, Doring}.

We show that the optimal value and solution of the approximate problem converge to the original problem, with probability one as the number of data points and resampling points increases to infinite (Theorems \ref{theo:convergence_tild_bbP_k} and~\ref{theo:almost_uniform_convergence}). 
A feasibility result (Theorem \ref{theo:finite_samples_feasibility}) is also demonstrated, proving that a feasible solution to the approximate problem is feasible to the original problem, with high probability, as the number of resampling points increases. 
We estimate the required resampling-point number to yield a lower bound and feasibility for the original problem.
Through a numerical example, we compare our proposed method against various existing approaches, demonstrating its superior ability to mitigate the bias typically associated with sample-based approximations, even in scenarios that do not directly account for conditional probabilities.

\vspace{-2mm}
%%%%%%%%%%%%%%%%%%%%%%%%%%%%%%%%%%%%%%%%%%%%%%%%%%%
%%%%%%%%%%%%%%%%%%%%%%%%%%%%%%%%%%%%%%%%%%%%%%%%%%%
\section{Problem statement}\label{sec:problem}

\subsection{System description}

Let $k\in\mathbb{N}$ be the discrete-time index. Define the system state and the vector of random disturbances of an autonomous system at $k$ by $\bm{\mathrm{x}}_k:=[x_{1,k}\ ...\ x_{n,k}]^\top\in\mathbb{R}^n$ and $\bm{\mathrm{w}}_k:=[w_{1,k}\ ...\ w_{s,k}]^\top\in\mathbb{R}^s,$ respectively. The state transition from time $k$ to time $k+1$ is described as
\begin{equation}
\label{eq:system}
    \bm{\mathrm{x}}_{k+1}=f(\bm{\mathrm{x}}_k,\bm{\mathrm{w}}_k),
\end{equation}
where $f:\mathbb{R}^n\times\mathbb{R}^s\rightarrow\mathbb{R}^n$ is a continuous nonlinear function, which is considered to be known. Although the nonlinear function is assumed to be known, the randomness of $\bm{\mathrm{w}}_k$ may capture some incomplete knowledge of the underlying dynamic model, particularly when the uncertainty may be dependent of system states and other variables.

For the random disturbances, the probability distribution of $\bm{\mathrm{w}}$ is considered to be conditioned on a \textit{contextual variable} ${\xi}_k:=[\bm{\mathrm{x}}_k^{\top},\bm{\mathrm{v}}_k^{\top}]^{\top},$ where $\bm{\mathrm{v}}_k\in\mathscr{V}\subseteq\mathbb{R}^m$ represents a time-varying environment variable. 
The term ``contextual" is adopted here since it generally represents the context at time $k$, which includes the state $\bm{\mathrm{x}}_k,$ and other time-variant environment factors $\bm{\mathrm{v}}_k$. We emphasize that these elements collectively define the context at each time $k$. Notably, ${\xi}_k$ excludes the input since this work focuses on autonomous systems.
However, by incorporating the input vector into $\bm{\mathrm{v}}_k$, the theoretical results of this paper can be directly extended to the control system with inputs.

\begin{myrema}
    The random disturbance in \eqref{eq:system} cannot be represented as $\bm{\mathrm{w}}_k({\xi}_k)$ as in \cite{Kohler}. This is because the mapping from ${\xi}_k$ to $\bm{\mathrm{w}}_k$ is not a function, as $\bm{\mathrm{w}}_k$ may not take on a single, identical value. Instead, the probability distribution of $\bm{\mathrm{w}}_k$ depends on ${\xi}_k$, and this distribution can potentially be multimodal.
\end{myrema}

Let $\bm{\mathrm{W}}$ and $\bm{\xi}$ be random disturbance and random contextual variable, respectively. $\bm{\mathrm{w}}$ is a realization of $\bm{\mathrm{W}}$ while ${\xi}$ is a realization of $\bm{\xi}$. Furthermore, their realizations at time $k$ are denoted by $\bm{\mathrm{w}}_k$ and ${\xi}_k$, respectively. Let $\Omega$ be a sample space and $\mathcal{F}$ be the $\sigma$-algebra of subsets of $\Omega$. With a probability measure $\mu_{\mathsf{o}}$ defined on Borel space $\left(\Omega,\mathcal{F}\right)$, a probability space $\left(\Omega,\mathcal{F},\mu_{\mathsf{o}}\right)$ is formed. Define an augmented vector $\bm{\delta}:=[\bm{\xi}^{\top},\bm{\mathrm{W}}^{\top}]^{\top}$, which has a support denoted by $\bm{\Delta}:=\bm{\Xi}\times\mathcal{W}$, where $\bm{\Xi}\subseteq\mathbb{R}^{n+m+1}$ and $\mathcal{W}\subseteq\mathbb{R}^q$. The vector $\bm{\delta}:\Omega\rightarrow\bm{\Delta}$ is a random variable and $ \delta := [{\xi}^{\top},\bm{\mathrm{w}}^{\top}]^{\top} \in\bm{\Delta}$ signifies a specific realization of $\bm{\delta}$ within $\bm{\Delta}$.

Let $\bm{\Delta}_{\mathsf{s}}\subseteq\bm{\Delta}$ be a subset of $\bm{\Delta}$. Given a continuous probability density function $p(\delta)$ with the support $\bm{\Delta}$, the probability that the random variable $\bm{\delta}$ resides within the subset $\bm{\Delta}_{\mathsf{s}}$ can be expressed as
\begin{equation*}
    % \label{eq:def_Pr}
   \mathsf{Pr}\{\bm{\delta}\in\bm{\Delta}_{\mathsf{s}}\}:=\int_{\bm{\Delta}_{\mathsf{s}}}p(\delta)\mathsf{d}\delta.
\end{equation*}

Recall that the random variable $\bm{\delta}$ comprises two components $\bm{\xi}$ and $\bm{\mathrm{W}}$. $\xi \in \bm{\Xi}$ and $\bm{\mathrm{w}} \in \mathcal{W}$ represent the realizations of $\bm{\xi}$ and $\bm{\mathrm{W}}$ within their regions, $\bm{\Xi}$ and $\mathcal{W}$, respectively. Suppose that the probability density function $p(\delta)$ is a joint density written by $p(\xi,\bm{\mathrm{w}})$ with $\delta=[{\xi}^{\top},\bm{\mathrm{w}}^{\top}]^{\top}$. Then, both $\bm{\xi}$ and $\bm{\mathrm{W}}$ are absolutely continuous with probability densities written by \cite{Capinski}  
\begin{align*}
    p_{\bm{\xi}}(\xi):=\int_{\mathcal{W}}p\left(\xi,\bm{\mathrm{w}}\right)\mathsf{d}\bm{\mathrm{w}},\  p_{\bm{\mathrm{W}}}(\bm{\mathrm{w}}):=\int_{\bm{\Xi}}p\left(\xi,\bm{\mathrm{w}}\right)\mathsf{d}\xi.
\end{align*}

Consider that $p_{\bm{\xi}}(\xi)>0$ and $p\left(\xi,\bm{\mathrm{w}}\right)>0$ hold for any $\xi\in\bm{\Xi}$ and any $\left(\xi,\bm{\mathrm{w}}\right)\in\bm{\Xi}\times\mathcal{W}$. We can define a conditional probability density by 
\begin{equation}
    \label{eq:def_condition}
    p_{\bm{\mathrm{W}}}^{\mathsf{c}}(\bm{\mathrm{w}}|\bm{\xi}=\xi):=\frac{p\left(\xi,\bm{\mathrm{w}}\right)}{p_{\bm{\xi}}(\xi)}.
\end{equation}
For notation simplicity, we omit ``$\bm{\xi}=$" in the conditional probability density \eqref{eq:def_condition} in the remainder of the paper. Additionally, we also define the conditional probability of having $\bm{\mathrm{W}}\in\mathcal{W}_{\mathsf{s}}\subseteq\mathcal{W}$ when $\bm{\xi}=\xi$ by 
\begin{equation}
    \label{eq:def_condition_probability}
    \mathsf{Pr}\left\{\bm{\mathrm{W}}\in\mathcal{W}_{\mathsf{s}}|\xi\right\}:=\int_{\mathcal{W}_{\mathsf{s}}} p_{\bm{\mathrm{W}}}^{\mathsf{c}} (\bm{\mathrm{w}}|\xi) \mathsf{d}\bm{\mathrm{w}} .
\end{equation}

\subsection{Problem formulation}

Let $\bm{\mathrm{e}}(\bm{\mathrm{x}}_k):=[1\ x_{1,k}\ ...\ x_{n,k}\ ...\ x_{1,k}^d\ ...\ x_{n,k}^d]^\top$
be a vector of monomials of degree up to $d>0$\footnote{For example, when $d=2,\ n=2$,  $\bm{\mathrm{e}}(\bm{\mathrm{x}}_k):=[1\ x_{1,k}\ x_{2,k}\ x_{1,k}x_{2,k}\ x_{1,k}^2\ x_{2,k}^2]^\top$.}. 
The polynomial function parameterized by $\bm{\theta}_k\in\mathbb{R}^{n_{\bm{\theta}}}$ as the probabilistic reachable set at time $k$ can be defined as
\begin{equation}\label{eq:poly_def}
q(\bm{\mathrm{x}}_k,\bm{\theta}_k):=\bm{\mathrm{e}}^\top(\bm{\mathrm{x}}_k)M(\bm{\theta}_k)\bm{\mathrm{e}}(\bm{\mathrm{x}}_k),
\end{equation}
where $n_{\bm{\theta}}$ is the dimension of the parameter vector~$\bm{\theta}_k,\ \forall k\in\mathbb{N}_+$ and $M(\bm{\theta}_k)$ is a symmetric gram matrix, which is positive semidefinite. 
Here, the parameter vector $\bm{\theta}_k$ including all the parameters in $M(\bm{\theta}_k).$ 
The degree of $q(\bm{\mathrm{x}}_k,\bm{\theta}_k)$ is $2d$. 
To proceed further, we restrict the range of $q(\bm{\mathrm{x}}_k,\bm{\theta}_k)$ to be non-negative, which yields $q(\bm{\mathrm{x}}_k,\bm{\theta}_k)$ to be polynomial sum-of-squares (SoS) \cite{Kozhasov}. 
For each time $k\in\mathbb{N}_+$, given a parameter vector $\bm{\theta}_k$ and the degree $d$, a polynomial sublevel set of $\bm{\mathrm{x}}_k$ as a probabilistic reachable set can be defined by
\begin{equation}
    \label{eq:polynomial_sublevel_set}
    \mathcal{V}\left(\bm{\theta}_k,d\right):=\{\bm{\mathrm{x}}_k\in\mathcal{X}_k:q(\bm{\mathrm{x}}_k,\bm{\theta}_k)\leq 1\},
\end{equation}
where $q(\bm{\mathrm{x}}_k,\bm{\theta}_k)$ is defined in \eqref{eq:poly_def}.

We are concerned with the probability that $\bm{\mathrm{x}}_k$ lies within the set $\mathcal{V}\left(\bm{\theta}_k,d\right)$, treated as a conditional probability given the initial state $\bm{\mathrm{x}}_0$ and consequence of environment variables $\{\bm{\mathrm{v}}_j\}_{j=1}^{k-1}$. For these dependent variables, let us define an augmented vector
\begin{equation}\label{eq:rho_k-1}
    \rho_{k-1}:=[\bm{\mathrm{x}}_0^{\top},\bm{\mathrm{v}}_0^{\top},...,\bm{\mathrm{v}}_{k-1}^{\top}]^{\top} \in \mathscr{P}_{k-1},
\end{equation}
where $\mathscr{P}_{k-1}:=\bm{\Xi}\times\mathscr{V}^{k-1}$.

\begin{myrema}
    To determine the probabilistic reachable set $\mathcal{V}\left(\bm{\theta}_k,d\right)$ of the state $\bm{\mathrm{x}}_k$ at each time $k$, it is necessary to have knowledge of $\rho_{k-1}$ as defined in \eqref{eq:rho_k-1}. Specifically, this requires the initial state $\bm{\mathrm{x}}_0$, which is given at time $k=0$, and the environment variables $\{\bm{\mathrm{v}}_j\}_{j=1}^{k-1}$, which must be known at time $k$.
\end{myrema}

The probability conditioned on $\rho_{k-1}$ and $\bm{\theta}_k$ is defined as
\begin{equation}\label{eq:def_contextual_P_k}
    \mathbb{P}_k(\bm{\theta}_k,\rho_{k-1}):=\int_{\mathcal{X}} \mathbb{I}_{1}(q(\bm{\mathrm{x}}_k,\bm{\theta}_k)) p_{\bm{\mathrm{X}}_k}^{\mathsf{c}}(\bm{\mathrm{x}}_k|\rho_{k-1}) \mathsf{d}\bm{\mathrm{x}}_k,
\end{equation}
where $\mathbb{I}_{1}(z)$ is an indicator function. If $z\leq 1$, then $\mathbb{I}_{1}(z)=1$. Otherwise, $\mathbb{I}_{1}(z)=0$. 

This work aims to identify the tightest set $\mathcal{V}\left(\bm{\theta}_k,d\right)$ such that $
\mathbb{P}_k(\bm{\theta}_k,\rho_{k-1})\geq 1-\alpha.$ 
Section 2.3 of \cite{Dabbene:2017} shows that the volume is a convex function of the defined polynomial when the polynomial is homogeneous, which is also introduced in \cite{Lasserre:2015, Morozov}.

A heuristic objective function has been used in~\cite{Magnani}. 
It is elucidated that maximizing the determinant of $M^{-1}(\bm{\theta}_k)$ serves to augment the curvature of $q(\bm{\mathrm{x}}_k,\bm{\theta}_k)$ across all possible directions. 
Consequently, minimizing the volume of $ \mathcal{V}\left(\bm{\theta}_k,d\right)$ effectively involves maximizing the determinant of $M^{-1}(\bm{\theta}_k)$. 
Given $\rho_{k-1}$, the addressed problem is formulated as
\begin{align} 
    &\min_{\bm{\theta}_k\in\Theta} \,\, J(\bm{\theta}_k,\rho_{k-1}):=\log\mathsf{det}\ M^{-1}(\bm{\theta}_k) \tag{$\mathcal{P}_{\alpha}^k(\rho_{k-1})$}, \label{eq:prob_orig}\\
    &s.t.\quad \bm{\mathrm{x}}_{j} = f(\bm{\mathrm{x}}_{j-1},\bm{\mathrm{w}}_{j-1}), \; j = 1,\ldots, k, \nonumber\\ 
    &\quad\quad\ \mathbb{P}_k(\bm{\theta}_k,\rho_{k-1}) \geq 1-\alpha. \label{eq:contextual_chance_constraint_V} 
\end{align}
Let $\Theta_{k,\alpha}(\rho_{k-1})$ be the feasibility region of problem \ref{eq:prob_orig} and $J^*_{k,\alpha}(\rho_{k-1}):=\min\{J(\bm{\theta}_k,\rho_{k-1}):\bm{\theta}_k\in\Theta_{k,\alpha}(\rho_{k-1})\}$
be the optimal objective value. Furthermore, the optimal solution set of problem $\mathcal{P}_{\alpha}^k(\rho_{k-1})$ is defined by $U_{k,\alpha}(\rho_{k-1}):=\{\bm{\theta}_k\in\Theta_{k,\alpha}(\rho_{k-1}): J(\bm{\theta}_k,\rho_{k-1})=J^*_{k,\alpha}(\rho_{k-1})\}.$ Therefore, $\bm{\theta}^*_{k,\alpha}(\rho_{k-1})\in U_{k,\alpha}(\rho_{k-1})$ is an optimal solution of problem \ref{eq:prob_orig}. 
The set $\mathcal{V}\left(\bm{\theta}^*_{k,\alpha}(\rho_{k-1}),d\right)$ specified by $\bm{\theta}^*_{k,\alpha}(\rho_{k-1})$ is called a \textit{minimal $(1-\alpha)$-reachable polynomial sublevel set of $\bm{\mathrm{x}}_k$ conditioned on $\rho_{k-1}.$} 
If problem \ref{eq:prob_orig} is solved efficiently, the minimum-volume polynomial sublevel set with chance constraints in \eqref{eq:contextual_chance_constraint_V} can be obtained.

\subsection{Challenging issue and the goals of the paper}

The existence of the chance constraint \eqref{eq:contextual_chance_constraint_V} introduces significant complexity to solving problem \ref{eq:prob_orig}. 
It is crucial to notice that the constraint \eqref{eq:contextual_chance_constraint_V} is not a general chance constraint, which involves random variables whose probability distribution is independent of other variables. 
Constraint \eqref{eq:contextual_chance_constraint_V} involves a conditional probability, and therefore \eqref{eq:contextual_chance_constraint_V} can also be called a \emph{contextual chance constraint} \cite{Rahimian}. 
An optimization problem incorporating contextual chance constraints is at least as difficult as one with general chance constraints. 
To effectively solve problem \ref{eq:prob_orig}, an understanding of the conditional probability of $\bm{\mathrm{X}}_k$ given $\rho_{k-1}$ is imperative. 

Optimization problems with chance constraints can be addressed using \emph{sample average approximation} (SAA) methods \cite{Luedtke, Pena} or \emph{scenario approach} (SA) \cite{Calafiore:2006, Campi:2008, Campi:2011}. 
These methods are collectively referred to as \textit{traditional sample-based methods} here.
Traditional sample-based methods are distribution-free, i.e. they do not require explicit probability distribution. Instead, they leverage a dataset of available historical observations of the uncertainties to approximately solve the chance-constrained optimization problems. 
In this context, ``distribution-free" means that these methods can handle uncertainties with any probability distribution, as long as the collected data is available. 
However, this introduces a significant limitation when applying traditional sample-based methods to problems involving contextual chance constraints, as the dataset must be sampled from the underlying probability distribution. 
In the following, we highlight this limitation by analyzing the statistical properties of the dataset obtained in the addressed problem with contextual chance constraints. 

In this work, the dataset of available historical observations of the system \eqref{eq:system} is represented by
\begin{equation}\label{eq:def_data_set}
    \mathcal{D}_N:=\left\{\xi^{(i)}_{\mathsf{s}},\bm{\mathrm{w}}^{(i)}_{\mathsf{s}}\right\}_{i=1}^{N},
\end{equation}
where disturbance samples $\bm{\mathrm{w}}^{(i)}_{\mathsf{s}},\ i=1,...,N$ are extracted according to the conditional probability distribution $p^{\mathsf{c}}_{\bm{\mathrm{W}}}(\bm{\mathrm{w}}|\xi_{\mathsf{s}}^{(i)})$, which is conditioned on~$\xi_{\mathsf{s}}^{(i)}.$

Since these contextual variable samples $\xi_{\mathsf{s}}^{(i)}, \; i = 1,\ldots, N$ vary widely across the dataset, ignoring these samples can result in a sample set 
\begin{equation}
    \mathcal{W}_N=\left\{\bm{\mathrm{w}}_{\mathsf{s}}^{(i)}\right\}_{i=1}^N,
\end{equation}
where for a sufficiently large $N$, this set can be treated as if it were drawn from the marginal distribution  $p_{\bm{\mathrm{W}}}(\bm{\mathrm{w}})$.

However, at each time $k$, the random disturbances $\bm{\mathrm{w}}_j, \; j=0,...,k-1$ follow the conditional probability $p^{\mathsf{c}}_{\bm{\mathrm{W}}}(\bm{\mathrm{w}}|\xi_{j})$, which depends on the contextual variable $\xi_{j} $. This is fundamentally different from the marginal distribution $p_{\bm{\mathrm{W}}}(\bm{\mathrm{w}})$, where the dependence on $\xi_{j} $ is ignored. As a result, if traditional sample-based methods rely solely on the sample set $\mathcal{W}_N$, the solution obtained would assume the disturbance samples follow the marginal distribution $p_{\bm{\mathrm{W}}}(\bm{\mathrm{w}})$ and the contextual dependence is ignored.

\begin{myrema}
    The discrepancy between the conditional distribution $p^{\mathsf{c}}_{\bm{\mathrm{W}}}(\bm{\mathrm{w}}|\xi_{j})$ and the marginal distribution $p_{\bm{\mathrm{W}}}(\bm{\mathrm{w}})$ implies that directly applying traditional sample-based methods would not provide an accurate or consistent approximate solution to the original problem \ref{eq:prob_orig}. Furthermore, in some cases, the conditional distribution $p^{\mathsf{c}}_{\bm{\mathrm{W}}}(\bm{\mathrm{w}}|\xi_{j})$ may be unknown, which makes it impossible to sample from this distribution. In Section \ref{sec:proposed}, we propose a resampling method to address this issue.
\end{myrema}

To address the above-discussed challenge issue of lacking the samples extracted according to $p^{\mathsf{c}}_{\bm{\mathrm{W}}}(\bm{\mathrm{w}}|\xi_{j})$, the goals of this paper are 
\begin{itemize}
    \item to propose a resampling method that leverages historical observation dataset to develop an enhanced sample-based approximation of problem \ref{eq:prob_orig}. 
    \item to design an efficient algorithm to compute probabilistic reachable sets $\mathcal{V}\left(\bm{\theta}_k,d\right)$ of system \eqref{eq:system}.
    \item to prove that the probabilistic reachable sets $\mathcal{V}\left(\bm{\theta}_k,d\right)$ converge almost surely to the desired ones given by problem \ref{eq:prob_orig}.
    \item to investigate the probabilistic feasibility of the solution by the proposed algorithm.
\end{itemize}

%%%%%%%%%%%%%%%%%%%%%%%%%%%%%%%%%%%%%%%%%%%%%%%%%%%
%%%%%%%%%%%%%%%%%%%%%%%%%%%%%%%%%%%%%%%%%%%%%%%%%%%
\section{The proposed algorithm}\label{sec:proposed}

\subsection{Targeted approximation of the original problem}

Consider that contextual variables' samples and disturbances in historical observation set $\mathcal{D}_N$ as in \eqref{eq:def_data_set} are independent and identically distributed (i.i.d.), $(\delta^{(i)}_{\mathsf{s}},\bm{\mathrm{w}}^{(i)}_{\mathsf{s}})\sim p(\delta,\bm{\mathrm{w}})$. 
Given a dataset $\mathcal{D}_N$ in \eqref{eq:def_data_set} and an initial contextual variable $\rho_{k-1}$, we propose a resampling-based method to approximate problem \ref{eq:prob_orig}. 

We first introduce an assumption on $q(\bm{\mathrm{x}}_k,\bm{\theta}_k)$ for the continuity analysis of $\mathbb{P}_k(\bm{\theta}_k,\rho_{k-1})$ in \eqref{eq:def_contextual_P_k}. 
Let $\mathsf{cl}\{Z\}$ be the closure of a set $Z$ and 
$\mathsf{supp}\ p_{\bm{\mathrm{X}}_k}^{\mathsf{c}}:=\mathsf{cl}\{\bm{\mathrm{x}}_k\in\mathcal{X}:p_{\bm{\mathrm{X}}_k}^{\mathsf{c}}(\bm{\mathrm{x}}_k|\rho_{k-1})>0\}.$ 
For each $\bm{\theta}_k\in\Theta\subseteq\mathbb{R}^{n_{\bm{\theta}}}$, let us define the set $\mathcal{X}^{\mathsf{supp}}_k(\bm{\theta}_k):=\left\{\bm{\mathrm{x}}_k\in\mathsf{supp}\ p_{\bm{\mathrm{X}}_k}^{\mathsf{c}}: q(\bm{\mathrm{x}}_k,\bm{\theta}_k)=1\right\}.$

\begin{myassump}\label{assump:Pbb_prob_measure}
    For each $\rho_{k-1}$ and $\bm{\theta}_k,\ k$, it holds
    \begin{equation}
           \mathsf{Pr}\left\{\mathcal{X}^{\mathsf{supp}}_k(\bm{\theta}_k)\right\}:=\int_{\mathcal{X}^{\mathsf{supp}}_k(\bm{\theta}_k)}p_{\bm{\mathrm{X}}_k}^{\mathsf{c}}(\bm{\mathrm{x}}_k|\rho_{k-1})\mathsf{d}\bm{\mathrm{x}}_k=0.
    \end{equation}
\end{myassump}

Assumption \ref{assump:Pbb_prob_measure} implies the continuity of $\mathbb{P}_k(\bm{\theta}_k,\rho_{k-1})$ \cite{Kibzun}, which guarantees  that the feasible set $\Theta_{k,\alpha}(\rho_{k-1})$ and optimal solution set $U_{k,\alpha}(\rho_{k-1})$ of problem \ref{eq:prob_orig} are measurable sets for all $\alpha\in[0,1],\ \rho_{k-1},\ k\in\mathbb{N}_+$.

In this paper, we propose a resampling algorithm based on the dataset $\mathcal{D}_N$ to generate $N_\mathsf{r}$ predicted state scenarios at any time $k$. The $i-$th predicted scenario at step $k$ is denoted by $\tilde{\bm{\mathrm{x}}}^{(i)}_{k}$ and $N_r$ predicted state scenarios $\tilde{\bm{\mathrm{x}}}^{(i)}_{k}$, $i = 1,\ldots, N_r$ form the set $\widetilde{\mathcal{X}}_k^{N_\mathsf{r}}$ as denoted below:
\begin{align}
    \widetilde{\mathcal{X}}_k^{N_\mathsf{r}}&:=\{\tilde{\bm{\mathrm{x}}}_k^{(i)}\}_{i=1}^{N_{\mathsf{r}}}.\label{eq:def_set_pre_sce_x_k}
\end{align}

The details of generating $\widetilde{\mathcal{X}}_k^{N_\mathsf{r}}$ is explained in Section \ref{subsection:algorithm_generation}.

By the predicted scenario set $\widetilde{\mathcal{X}}_k^{N_\mathsf{r}}$ in \eqref{eq:def_set_pre_sce_x_k}, we formulate $\mathbb{P}_k(\bm{\theta}_k,\rho_{k-1})$'s sample-based approximation in \eqref{eq:def_contextual_P_k} as
\begin{equation}\label{eq:contextual_P_estimate}
    \widetilde{\mathbb{P}}_k(\bm{\theta}_k,\rho_{k-1},\mathcal{D}_N,\widetilde{\mathcal{X}}_k^{N_\mathsf{r}}):=\frac{1}{N_{\mathsf{r}}}\sum_{i=1}^{N_{\mathsf{r}}} \mathbb{I}_{1}(q(\tilde{\bm{\mathrm{x}}}^{(i)}_{k},\bm{\theta}_k)).
\end{equation}

Then, an approximate problem of problem \ref{eq:prob_orig} can be formulated as
\begin{align} 
    &\min_{\bm{\theta}_k\in\Theta} \,\, \log\mathsf{det}\ M^{-1}(\bm{\theta}_k) \tag{$\widetilde{\mathcal{P}}_{\alpha_{\mathsf{s}}}^k(\rho_{k-1},\mathcal{D}_N,\widetilde{\mathcal{X}}_k^{N_\mathsf{r}})$} \label{eq:prob_approximate}\\
    \mathsf{s.t.} & \quad \widetilde{\mathbb{P}}_k(\bm{\theta}_k,\rho_{k-1},\mathcal{D}_N,\widetilde{\mathcal{X}}_k^{N_\mathsf{r}})\geq 1-\alpha_{\mathsf{s}}. \label{eq:constraint_approximate}
\end{align}

Solving problem \ref{eq:prob_approximate} is still challenging since \eqref{eq:constraint_approximate} is a non-smooth constraint due to the indicator function used in \eqref{eq:contextual_P_estimate}. To address this issue, we adopt the algorithm in \cite{Pena} to solve it by using the smooth approximation of the indicator function.

Let $\widetilde{\Theta}_{k,\alpha_{\mathsf{s}}}(\rho_{k-1},\mathcal{D}_N,\widetilde{\mathcal{X}}_k^{N_\mathsf{r}})$ be the feasibility region of problem \ref{eq:prob_approximate} and 
\begin{align*}
    \widetilde{J}_{k,\alpha_{\mathsf{s}}}& (\rho_{k-1},\mathcal{D}_N,\widetilde{\mathcal{X}}_k^{N_\mathsf{r}}):=\nonumber\\
    & \min\{J(\bm{\theta}_k,\rho_{k-1}):
     \bm{\theta}_k\in\widetilde{\Theta}_{k,\alpha_{\mathsf{s}}}(\rho_{k-1},\mathcal{D}_N,\widetilde{\mathcal{X}}_k^{N_\mathsf{r}})\}  &\label{eq:opt_obj_value_approximate}
\end{align*}
be the optimal objective value. Furthermore, the optimal solution set of problem \ref{eq:prob_approximate} is defined by 
\begin{align*}
    % \label{eq:opt_sol_set_approximate}
    \widetilde{U}_{k,\alpha_{\mathsf{s}}}& (\rho_{k-1},\mathcal{D}_N,\widetilde{\mathcal{X}}_k^{N_\mathsf{r}}):=\{ \bm{\theta}_k\in\widetilde{\Theta}_{k,\alpha_{\mathsf{s}}}(\rho_{k-1},\mathcal{D}_N,\widetilde{\mathcal{X}}_k^{N_\mathsf{r}}): \nonumber \\
    & J(\bm{\theta}_k,\rho_{k-1})=\widetilde{J}_{k,\alpha_{\mathsf{s}}}(\rho_{k-1},\mathcal{D}_N,\widetilde{\mathcal{X}}_k^{N_\mathsf{r}})\}.
\end{align*}
The optimal solution is denoted by $\tilde{\bm{\theta}}_{k,\alpha_{\mathsf{s}}}(\rho_{k-1},\mathcal{D}_N,\widetilde{\mathcal{X}}_k^{N_\mathsf{r}}),$ with abbreviation as $\tilde{\bm{\theta}}_k$.
Notice that the predicted scenario set $\widetilde{\mathcal{X}}_k^{N_\mathsf{r}}$ can be obtained from $\bm{\mathrm{x}}_0$ and $\mathcal{D}_N$ by a designed algorithm. 
If $\rho_{k-1}$ is given, the optimal objective value $\widetilde{J}_{k,\alpha_{\mathsf{s}}}(\rho_{k-1},\mathcal{D}_N,\widetilde{\mathcal{X}}_k^{N_\mathsf{r}})$ and the optimal solution set $\widetilde{U}_{k,\alpha_{\mathsf{s}}}(\rho_{k-1},\mathcal{D}_N,\widetilde{\mathcal{X}}_k^{N_\mathsf{r}})$ are dependent on the dataset $\mathcal{D}_N$ and sample set $\widetilde{\mathcal{X}}_k^{N_\mathsf{r}}$. 
Since $\mathcal{D}_N$ and $\widetilde{\mathcal{X}}_k^{N_\mathsf{r}}$ are randomly generated, the optimal objective value $\widetilde{J}_{k,\alpha_{\mathsf{s}}}(\rho_{k-1},\mathcal{D}_N,\widetilde{\mathcal{X}}_k^{N_\mathsf{r}})$ is a random variable, and the optimal solution set $\widetilde{U}_{k,\alpha_{\mathsf{s}}}(\rho_{k-1},\mathcal{D}_N,\widetilde{\mathcal{X}}_k^{N_\mathsf{r}})$ is a random set. 

\begin{figure*}[t]
\centering
\includegraphics[width = \hsize]{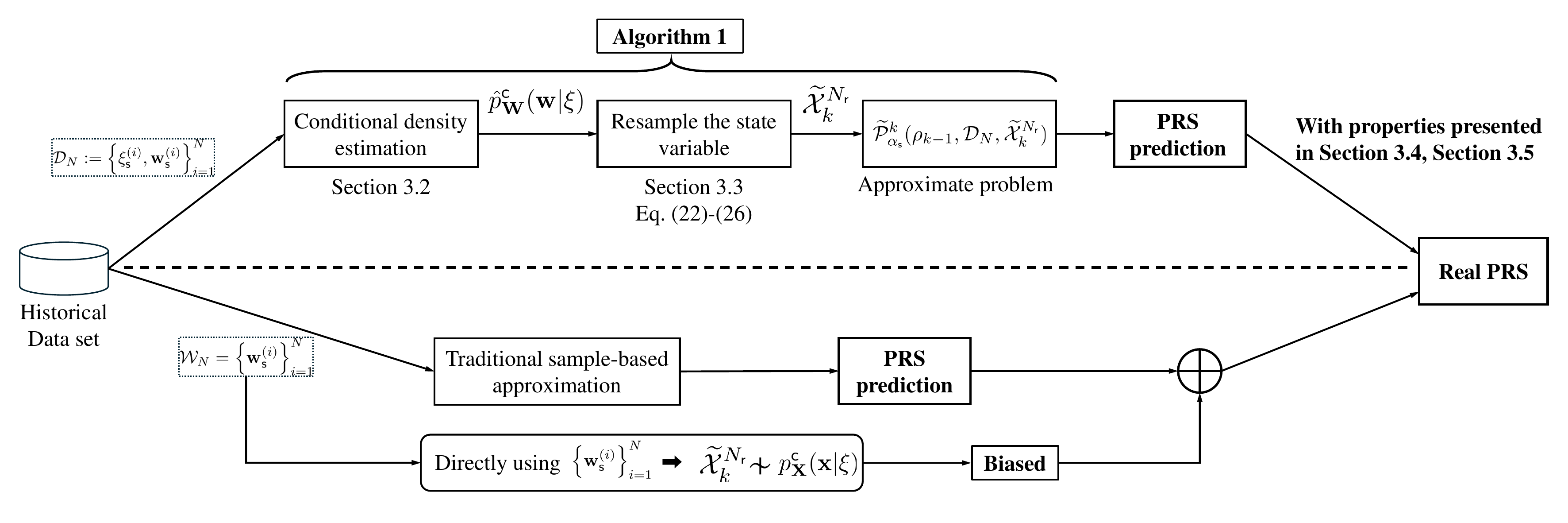}
\centering
\caption{The overall structure of the proposed algorithm compared to traditional sample-based methods (PRS: probabilistic reachable set).}
\label{fig:contribution_summary}
\end{figure*}

In the sequel, we aim to delve into the following tasks:
\begin{itemize}
    \item Utilizing the dataset $\mathcal{D}_N$ to achieve a conditional probability density estimation (Section \ref{subsection:cond_den_est});
    \item Designing resampling algorithm to generate samples by estimated probability density (Section \ref{subsection:algorithm_generation});
    \item Exploring whether the optimal objective value and the optimal solution set of the approximate problem \ref{eq:prob_approximate} converges to those of the original problem \ref{eq:prob_orig} as the number of sample goes to infinity (Section \ref{subsection:almost_uni_conv});
    \item Investigating if Problem \ref{eq:prob_approximate}'s optimal solution can be ensured within a bounded probability, given only finite samples available (Section \ref{subsection:prob_fea_fini_sam}).
\end{itemize}

The overall structure of the algorithm is illustrated in Fig.~\ref{fig:contribution_summary}, alongside a comparison with traditional sample-based methods. 
Traditional sample-based methods directly use the samples $\bm{\mathrm{w}}_{\mathsf{s}}^{(i)},\ i=1,...,N$, which cannot obtain the samples of the state at time $k$ that follow the conditional distribution $p^{\mathsf{c}}_{\bm{\mathrm{X}}}(\bm{\mathrm{x}}|\rho_{k-1})$.  As a result, the obtained probabilistic reachable set produced by traditional methods may deviate from the true reachable sets.

In subsection \ref{subsection:cond_den_est}, we introduce a method that leverages the dataset $\mathcal{D}_N$ to train a model that estimates the conditional probability density for any given $\xi_j$. Based on the estimated conditional density, the state $\widetilde{\mathcal{X}}_k^{N_{\mathsf{r}}}$ can be resampled. 
These resampled states are used to construct the approximate problem \ref{eq:prob_approximate}. 
The above procedures are the core of Algorithm \ref{alg:rsSAA}. In subsections \ref{subsection:almost_uni_conv} and \ref{subsection:prob_fea_fini_sam}, we give the theoretical results of the proposed method, almost uniform convergence and probabilistic feasibility. Specifically, the probabilistic reachable set obtained by the proposed method converges to the minimal $(1-\alpha)-$reachable polynomial sublevel set by problem \ref{eq:prob_orig} when $N,\ N_{\mathsf{r}}\rightarrow\infty$. Moreover, with a sufficiently large $N_{\mathsf{r}}$, this set can represent a $(1-\alpha)$-reachable polynomial sublevel set. 

\subsection{Conditional density estimation}
\label{subsection:cond_den_est}

We first introduce a conditional density estimation method, leveraging historical dataset $\mathcal{D}_N$. The conditional density estimation is pivotal for the design of our resampling algorithm. Inspired by \cite{Rahimian}, the weighted kernel density estimation can be applied to calculate the weights of samples. However, the weighted kernel density estimation is unreliable in high-dimensional problems \cite{Wolff}. This paper adapts the least-squares conditional density estimation method presented in \cite{Sugiyama_AISM, Kanamori, Sugiyama} to this problem. Based on the conditional density estimation, we compute the samples yielding the probability distribution of the states and disturbances. 

We use a function to denote the conditional probability density $p_{\bm{\mathrm{W}}}^{\mathsf{c}}(\bm{\mathrm{w}}|\xi)$ as
\begin{equation}
    \label{eq:def_density_ratio_function}
    \pi\left(\bm{\mathrm{x}},\xi\right):=p_{\bm{\mathrm{W}}}^{\mathsf{c}}(\bm{\mathrm{w}}|\xi)=\frac{p\left(\bm{\mathrm{x}},\xi\right)}{p_{\bm{\xi}}(\xi)}.
\end{equation}
$\pi\left(\xi,\bm{\mathrm{w}}\right)$ expresses the essential that the conditional probability density is a function of $\xi$ and $\bm{\mathrm{w}}$, which is called the \textit{density ratio function}. 

Let us define a basis function as
\begin{equation}
    \label{eq:def_basis_function}
    \bm{\phi}\left(\xi,\bm{\mathrm{w}}\right):=[\phi_1\left(\xi,\bm{\mathrm{w}}\right),...,\phi_b\left(\xi,\bm{\mathrm{w}}\right)]^\top,
\end{equation}
which is required to satisfy $\bm{\phi}\left(\xi,\bm{\mathrm{w}}\right)\geq \bm{0}^{b}$ for all $\left(\xi,\bm{\mathrm{w}}\right)\in\bm{\Xi}\times \mathcal{W}.$ With this basis function, we model the density ratio function $ \pi\left(\xi,\bm{\mathrm{w}}\right)$ defined in \eqref{eq:def_density_ratio_function} by a linear model as
\begin{equation}
    \label{eq:model_density_ratio_function}
    \hat{\pi}_{\bm{\beta}}\left(\xi,\bm{\mathrm{w}}\right):=\bm{\beta}^\top  \bm{\phi}\left(\xi,\bm{\mathrm{w}}\right),
\end{equation}
where $\bm{\beta}:=[\beta_1,...,\beta_b]^\top$ is the parameter vector to be learned from samples. Notice that we can adapt the model $\hat{\pi}_{\bm{\beta}}\left(\xi,\bm{\mathrm{w}}\right)$ to a kernel model by setting the basis function $\bm{\phi}\left(\xi,\bm{\mathrm{w}}\right)$ dependent on all samples in $\mathcal{{D}}_N$ and choosing the basis function number $b=N$. 

For a given basis function $\bm{\phi}\left(\xi,\bm{\mathrm{w}}\right)$ and a dataset $\mathcal{D}_N,$ the objective is to optimize the linear coefficient $\bm{\beta}$ such that the density ratio function $\hat{\pi}_{\bm{\beta}}\left(\xi,\bm{\mathrm{w}}\right)$ closely approximates the conditional probability density $\pi\left(\xi,\bm{\mathrm{w}}\right).$  To achieve this, a loss function is required for the optimization. Moreover, it requires that the estimated density $\hat{\pi}_{\bm{\beta}}\left(\xi,\bm{\mathrm{w}}\right)$ converges to the conditional probability density $\pi\left(\xi,\bm{\mathrm{w}}\right)$, as the sample number approaches infinity.

With a dataset $\mathcal{D}_N$, let us define a sample-based loss function as
\begin{equation}\label{eq:loss function}
    \widehat{\mathcal{L}}_{\lambda}(\bm{\beta}):=\bm{\beta}^\top \widehat{\bm{\mathrm{H}}}\bm{\beta} - 2\hat{\bm{\mathrm{h}}}^\top\bm{\beta} + \lambda \bm{\beta}^\top\bm{\beta},
\end{equation}
where
\begin{subequations}
    \begin{align}
        \widehat{\bm{\mathrm{H}}} &:=\frac{1}{N}\sum_{i=1}^N \int_{\mathcal{W}} \bm{\phi}\left(\xi_{\mathsf{s}}^{(i)},\bm{\mathrm{w}}\right) \bm{\phi}\left(\xi_{\mathsf{s}}^{(i)},\bm{\mathrm{w}}\right)^\top \mathsf{d}\bm{\mathrm{w}},\label{eq:def_hat_H}\\  
        \hat{\bm{\mathrm{h}}}^\top &:=\frac{1}{N}\sum_{i=1}^N \bm{\phi}\left(\xi_{\mathsf{s}}^{(i)},\bm{\mathrm{w}}_{\mathsf{s}}^{(i)}\right).\label{eq:def_hat_h}
    \end{align}
\end{subequations}

Based on the defined loss function in \eqref{eq:loss function}, the analytical solution of the learned parameter can be formulated as
\begin{equation}
    \label{eq:beta_opt}
    \tilde{\bm{\beta}}:=\arg\min_{\bm{\beta}\in\mathbb{R}^b}  \widehat{\mathcal{L}}_{\lambda}(\bm{\beta})=\left(\widehat{\bm{\mathrm{H}}}+\lambda \bm{\mathrm{I}}_{b} \right)^{-1} \hat{\bm{\mathrm{h}}},
\end{equation}
where $\bm{\mathrm{I}}_b$ is the $b-$dimensional identity matrix. With this learned parameter $\tilde{\bm{\beta}}$ in \eqref{eq:beta_opt}, the \emph{Least-Squares Conditional Density Estimation} (LS-CDE) is given as \cite{Sugiyama}
\begin{equation}
    \label{eq:LS_CDE}
    \hat{p}^{\mathsf{c}}_{\bm{\mathrm{W}}}(\bm{\mathrm{w}}|\xi):=\frac{\tilde{\bm{\beta}}^\top \bm{\phi}\left(\xi,\bm{\mathrm{w}}\right)}{\int_{\mathcal{W}}\tilde{\bm{\beta}}^\top \bm{\phi}\left(\xi,\bm{\mathrm{w}}\right)\mathsf{d}\bm{\mathrm{w}}}.
\end{equation}

The convergence of $\hat{p}^{\mathsf{c}}_{\bm{\mathrm{W}}}(\bm{\mathrm{w}}|\xi)$ is given in \cite[Theorem 1]{Sugiyama}. 
We briefly review the convergence analysis for the LS-CDE. Note that the learned parameter $\tilde{\bm{\beta}}$ depends on the dataset $\mathcal{D}_N$ and the coefficient $\lambda$. 
To link with the number of samples $N$, we re-denote this coefficient by $\lambda_N$. 
Let $\mu_{\bm{\xi}}$ be the probability measure on $\bm{\Xi}$ associated with $p_{\bm{\xi}}(\cdot)$ and $\mu_{\bm{\mathrm{W}}}$ be finite Lebesgue measure $\mathcal{W}$. 
We use $\mu_{\bm{\xi}} \times \mu_{\bm{\mathrm{W}}}$ to denote a product measure of $\mu_{\bm{\xi}}$ and $\mu_{\bm{\mathrm{W}}}$. \cite[Theorem 1]{Sugiyama} is summarized as the following lemma.

\begin{mylemma}
    \label{lemma:convergence_LS_CDE}
    As $N\rightarrow\infty, $ if $\lambda_N\rightarrow 0$ with $\lambda_N^{-1}=o(N^{2/(2+\gamma)}),\ \gamma\in(0,2),$ then
    \begin{equation}
        \label{eq:convergence_LS_CDE_pi}
        \|\hat{\pi}_{\bm{\beta}}-\pi\|_{2}=\mathcal{O}_p(\lambda_N^{1/2}),
    \end{equation}
    where $\|\cdot\|_{2}$ is the $L_2\left(\mu_{\bm{\xi}} \times \mu_{\bm{\mathrm{W}}}\right)$-norm, and $\mathcal{O}_p$ is the asymptotic order in probability. 
\end{mylemma}

Based on the LS-CDE $\hat{p}^{\mathsf{c}}_{\bm{\mathrm{W}}}(\bm{\mathrm{w}}|\xi)$, an approximation of the probability $\mathbb{P}_k(\bm{\theta}_k,\rho_{k-1})$ in \eqref{eq:def_contextual_P_k} can be formulated as
\begin{equation}
    \label{eq:contextual_P_approximate_transit}
    \widehat{\mathbb{P}}_k(\bm{\theta}_k,\bm{\mathrm{x}}_0,\mathcal{D}_N):=\int_{\mathcal{X}} \mathbb{I}_{1}(q(\bm{\mathrm{x}}_k,\bm{\theta}_k)) \hat{p}^{\mathsf{c}}_{\bm{\mathrm{X}}_k}(\bm{\mathrm{x}}_k|\xi_0) \mathsf{d}\bm{\mathrm{x}}_k,
\end{equation}
where $\hat{p}^{\mathsf{c}}_{\bm{\mathrm{X}}_k}(\bm{\mathrm{x}}_k|\xi_0)$ is the estimated probability density of $\bm{\mathrm{x}}_k$ conditioned on $\xi_0$ when using the LS-CDE $\hat{p}^{\mathsf{c}}_{\bm{\mathrm{W}}}(\bm{\mathrm{w}}|\xi)$ instead of the actual condition density $p^{\mathsf{c}}_{\bm{\mathrm{W}}}(\bm{\mathrm{w}}|\xi) $ (which is likely unknown in practice). Note that we can choose the basis function $\bm{\phi}(\cdot)$ in \eqref{eq:def_basis_function}, e.g. a Gaussian function, to ensure $\hat{p}^{\mathsf{c}}_{\bm{\mathrm{X}}_k}(\bm{\mathrm{x}}_k|\xi_0)$ satisfying Assumption \ref{assump:Pbb_prob_measure}. 

\subsection{Resampling-based probabilistic reachable sets}
\label{subsection:algorithm_generation}

Leveraging the estimated conditional probability density, we next propose a recursive algorithm designed to generate the predicted scenario set at each time $k$. Then, this dataset is used to solve the approximate problem, thereby computing probabilistic reachable sets of stochastic states. 

Starting from $\bm{\mathrm{x}}_0$ with the initial contextual variable $\xi_0$, we first extracts $N_{\mathsf{r}}$ samples from $\mathcal{W}$ according to the LS-CDE $\hat{p}^{\mathsf{c}}_{\bm{\mathrm{W}}}(\bm{\mathrm{w}}|\xi)$ as follows:
\begin{equation}
    \label{eq:resample_w_0}
    \widetilde{\bm{\mathrm{W}}}_0:=\left\{\tilde{\bm{\mathrm{w}}}_0^{(i)}\right\}_{i=1}^{N_\mathsf{r}},\ \tilde{\bm{\mathrm{w}}}_0^{(i)}\sim\hat{p}^{\mathsf{c}}_{\bm{\mathrm{W}}}(\bm{\mathrm{w}}|\xi_0).  
\end{equation}

The predicted scenarios at $k=1$ are then computed by
\begin{equation}
    \label{eq:compute_predicted_scenario_1}
    \tilde{\bm{\mathrm{x}}}_1^{(i)}=f(\bm{\mathrm{x}}_0,\tilde{\bm{\mathrm{w}}}^{(i)}_{0}),\ i=1,...,N_{\mathsf{r}},
\end{equation}
which gives the predicted scenario set of $\tilde{\bm{\mathrm{x}}}_1^{(i)}$ as
\begin{equation}
    \label{eq:def_set_pre_sce_x_1}
    \widetilde{\mathcal{X}}_1:=\{\tilde{\bm{\mathrm{x}}}_1^{(i)}\}_{i=1}^{N_\mathsf{r}}.
\end{equation}
For time $j\geq 1$, define $\tilde{\xi}^{(i)}_{j}:=\{\tilde{\bm{\mathrm{x}}}_{j}^{(i)},\bm{\mathrm{v}}_{j},t_{j}\}$, we extract $N_{\mathsf{r}}$ samples from $\mathcal{W}$ by
\begin{equation}
    \label{eq:resample_w_k_1}
    \widetilde{\mathcal{W}}_{j}:=\{\tilde{\bm{\mathrm{w}}}_{j}^{(i)}\}_{i=1}^{N_\mathsf{r}},\ \tilde{\bm{\mathrm{w}}}_{j}^{(i)}\sim\hat{p}^{\mathsf{c}}_{\bm{\mathrm{W}}}(\bm{\mathrm{w}}|\tilde{\xi}^{(i)}_{j}),
\end{equation}
and the predicted scenarios at $j+1$ are consequently computed from $\widetilde{\mathcal{W}}_{j}$ by
\begin{equation}
    \label{eq:compute_predicted_scenario_k_rs}
    \tilde{\bm{\mathrm{x}}}_{j+1}^{(i)}=f(\tilde{\bm{\mathrm{x}}}^{(i)}_{j},\tilde{\bm{\mathrm{w}}}^{(i)}_{j}),\ i=1,...,N_{\mathsf{r}},
\end{equation}
which forms the predicted scenario set $\widetilde{\mathcal{X}}_{j+1}^{N_\mathsf{r}}$ as in \eqref{eq:def_set_pre_sce_x_k} for $j \geq 1$. The resampling-based algorithm to obtain probabilistic reachable sets is summarized in Algorithm~\ref{alg:rsSAA}.

\begin{algorithm}[t]
    \caption{Resampling-based probabilistic reachable sets.}\label{alg:rsSAA}
    \begin{algorithmic}
        \State {\textbf{Inputs:}} Data set $\mathcal{D}_N$, predictive horizon $k$, variable $\rho_{k-1}:=\{\bm{\mathrm{x}}_0,\bm{\mathrm{v}}_0,t_0,...,\bm{\mathrm{v}}_{k-1},t_{k-1}\}$, probability level $\alpha_{\mathsf{s}}$;
        \State Establish the basis function $\bm{\mathrm{\phi}}(\xi,\bm{\mathrm{w}})$ in \eqref{eq:def_basis_function} and compute the parameter $\tilde{\bm{\beta}}$ by \eqref{eq:beta_opt};
        \State $j \leftarrow 0$;
        \While{$j < k$}
        \If{$j=0$}
            \State Sample $\widetilde{\bm{\mathrm{W}}}_{0}$ by \eqref{eq:resample_w_0};
            \State Compute the predicted scenario $\tilde{\bm{\mathrm{x}}}_{1}$ by \eqref{eq:compute_predicted_scenario_1};
            \State Obtain the predicted scenario set $\widetilde{\mathcal{X}}_{1}^{N_\mathsf{r}}$;
        \Else
            \State Sample $\widetilde{\bm{\mathrm{W}}}_{j}$ by \eqref{eq:resample_w_k_1} and the predicted scenario $\tilde{\bm{\mathrm{x}}}_{j+1}$ by \eqref{eq:compute_predicted_scenario_k_rs};
            \State Obtain the predicted scenario set $\widetilde{\mathcal{X}}_{j+1}^{N_\mathsf{r}}$;
        \EndIf
        \State $j \leftarrow j+1$;
        \EndWhile
        \State Solve problem \ref{eq:prob_approximate} to obtain $\tilde{\bm{\theta}}_k$;
        \State {\textbf{Output:}} $\tilde{\bm{\theta}}_k$
    \end{algorithmic}
\end{algorithm}

\subsection{Almost uniform convergence}\label{subsection:almost_uni_conv}

Standard approaches for proving convergence results in traditional sample-based methods involve first demonstrating that the constraint function of the sample-based approximation converges to the original constraint function with probability 1, followed by addressing the convergence of the optimal value and solution. We refer the reader to \cite{Pena, Shen2023_JOTA}, which provides both inspiration and a point of comparison for the results derived in this paper.

In contrast to noncontextual cases, we here consider data points resampled according to a probability distribution that depends on the contextual variable $\xi_j, \; j=0,1,...,k-1$. Additionally, the error analysis of the estimated conditional density must be considered. As a result, techniques used to prove the convergence of the constraint function in noncontextual cases cannot be directly applied or extended to contextual scenarios.

The theoretical results below proceed in two major steps. First, we establish the convergence of the resampling-based approximate constraint at time $k=1$ over any compact set for $(\bm{\theta}_k, \rho_{k-1})$, by analyzing both the sample-based error and the error in conditional density estimation. Next, we demonstrate that if convergence holds at time~$k$, the residual of the approximate constraint at time~$k+1$ will also converge. Using mathematical induction, we can then prove the convergence of the approximate constraint for all times~$k$.

It is important to note that in non-contextual cases, the convergence of the approximate constraint does not require considering errors from conditional density estimation, nor does it require mathematical induction since the probability density remains identical for each $k$.

Next, we present our theoretical results on almost uniform convergence. The solutions of \ref{eq:prob_approximate} might deviate from the solutions of \ref{eq:prob_orig} due to that
\begin{itemize}
    \item[(a)] The sample set $\widetilde{\mathcal{X}}_k^{N_{\mathsf{r}}}$ is extracted based on the LS-CDE $ \hat{p}^{\mathsf{c}}_{\bm{\mathrm{W}}_k}(\bm{\mathrm{w}}_k|\xi_k)$ instead of the actual conditional probability density $p^{\mathsf{c}}_{\bm{\mathrm{W}}_k}(\bm{\mathrm{w}}_k|\xi_k)$ as it is unknown;
    \item[(b)] The constraint \eqref{eq:constraint_approximate} in the approximate problem \ref{eq:prob_approximate} is discrete approximation of the probability integral. 
\end{itemize}

In the following, we give two theoretical results of almost uniform convergence for problem \ref{eq:prob_approximate}. We first establish that the approximate constraint \eqref{eq:constraint_approximate} converges to the original constraint outlined in \eqref{eq:contextual_chance_constraint_V} with probability one as the number of samples approaches infinity. Subsequently, we demonstrate that the set of optimal solutions for the approximate problem \ref{eq:prob_approximate} also converges to a subset of the optimal solution set for the original problem \ref{eq:prob_orig} with probability one as the number of samples approaches infinity.

First, we present the almost uniform convergence of the approximate constraint \eqref{eq:constraint_approximate} in the following theorem.
\begin{mytheo}
    \label{theo:convergence_tild_bbP_k}
    Suppose that Assumption \ref{assump:Pbb_prob_measure} holds. Consider that $\lambda_{N}\rightarrow 0$ with $\lambda_N^{-1}=o(N^{2/(2+\gamma)}),\ \gamma\in(0,2)$ as $N\rightarrow\infty.$ Then, 
    \begin{equation}
        \label{eq:convergence_tild_bbP_k_k}
        \sup_{\substack{\bm{\theta}_k\in\Theta_{\mathsf{c}} \\ \rho_{k-1}\in\mathscr{P}_{k-1}^{\mathsf{c}}}}\left|\widetilde{\mathbb{P}}_k(\bm{\theta}_k,\rho_{k-1},\mathcal{D}_N,\widetilde{\mathcal{X}}_k^{N_\mathsf{r}})-\mathbb{P}_k(\bm{\theta}_k,\rho_{k-1})\right|\rightarrow 0,
    \end{equation}
    with probability one as $N, N_{\mathsf{r}}\rightarrow\infty$, where $\Theta_{\mathsf{c}}$ is any compact subset of $\Theta\subseteq\mathbb{R}^{n_{\bm{\theta}}}$, and for any $\epsilon>0$, there exists a subset $\mathscr{P}_{k-1}^{\mathsf{c}}$\footnote{Notice that we choose the same notation $\mathscr{P}_{k-1}^{\mathsf{c}}$ for different cases since it can be easily proved that we could always find a common part that can be extended to the one with probability measure sufficiently close to 1 due to the assumption that the probability density is positive everywhere.} of $\mathscr{P}_{k-1}$ such that $\mu_{\rho}(\mathscr{P}_{k-1}\setminus\mathscr{P}_{k-1}^{\mathsf{c}})<\epsilon$, where $\mu_{\rho}$ is an arbitrary probability measure on $\mathscr{P}_{k-1}$ associated with positive probability density everywhere.
\end{mytheo}

\begin{pf}
    The proof can be found in Appendix \ref{proof:theo_convergence_tild_bbP_k}.  \qed
\end{pf}

Theorem \ref{theo:convergence_tild_bbP_k} shows that the approximate constraint \eqref{eq:constraint_approximate} converges uniformly on any compact set of $\bm{\theta}_k$ and a subset of $\mathscr{P}_{k-1}$ with probability measure sufficiently close to 1. 

We next give the almost uniform convergence of the proposed resampling-based approximate method as summarized in Algorithm \ref{alg:rsSAA} for problem \ref{eq:prob_approximate}. 

\begin{mytheo}
    \label{theo:almost_uniform_convergence}
    Suppose that Assumption \ref{assump:Pbb_prob_measure} holds. Assume that $\Theta$ is compact and $\lambda_{N}\rightarrow 0$ with $\lambda_N^{-1}=o(n^{2/(2+\gamma)}),\ \gamma\in(0,2)$ as $N\rightarrow\infty.$ Then, for every $\epsilon>0$ and if $k<\infty$, there exists a compact subset $\mathscr{P}_{k-1}^{\mathsf{c}}\subset\mathscr{P}_{k-1}$ such that $\mu_{\rho}(\mathscr{P}_{k-1}\setminus\mathscr{P}_{k-1}^{\mathsf{c}})<\epsilon$, where $\mu_{\rho}$ is an arbitrary probability measure on $\mathscr{P}_{k-1}$ associated with positive probability density everywhere, 
    \begin{equation}
        \label{eq:almost_uniform_convergence_J}
        \sup_{\rho_{k-1}\in\mathscr{P}_{k-1}^{\mathsf{c}}}\left|\widetilde{J}_{k,\alpha}(\rho_{k-1},\mathcal{D}_N,\widetilde{\mathcal{X}}_k^{N_\mathsf{r}})- J^{*}_{k,\alpha}(\rho_{k-1})\right|\rightarrow 0,
    \end{equation}
    \begin{equation}
        \label{eq:almost_uniform_convergence_bbD}
        \sup_{\rho_{k-1}\in\mathscr{P}_{k-1}^{\mathsf{c}}}\mathbb{D}\left(\widetilde{U}_{k,\alpha}(\rho_{k-1},\mathcal{D}_N,\widetilde{\mathcal{X}}_k^{N_\mathsf{r}}),U_{k,\alpha}(\rho_{k-1})\right)\rightarrow 0,
    \end{equation}
    with probability 1 as $N,N_{\mathsf{r}}\rightarrow\infty$, where $$\mathbb{D}(Z_1,Z_2):=\sup_{z\in Z_1}\inf_{z'\in Z_2}\|z-z'\|$$ is the deviation of set $Z_1$ from set $Z_2$.
\end{mytheo}

\begin{pf}
    The proof can be found in Appendix \ref{proof:theo_almost_uniform_convergence}. \qed
\end{pf}

Theorem \ref{theo:almost_uniform_convergence} exhibits that the resampling-based approximate problem \ref{eq:prob_approximate}'s optimal solution set converges to a subset of the original problem \ref{eq:prob_orig} for $\rho_{k-1}$ on a set that could be a sufficiently large subset of $\mathscr{P}_{k-1}$. This conclusion implies that the reachable set obtained by solving the approximate problem \ref{eq:prob_approximate} converges to a desired minimal $(1-\alpha)-$reachable polynomial sublevel set with probability~1 for every $\rho_{k-1}$ in $\mathscr{P}_{k-1}$. The uniform convergence can be ensured if $\mathscr{P}_{k-1}$ is compact. 

\subsection{Probabilistic feasibility with finite samples}
\label{subsection:prob_fea_fini_sam}

Theorem \ref{theo:almost_uniform_convergence} demonstrates that the optimal solution of problem \ref{eq:prob_orig} can be achieved by solving problem \ref{eq:prob_approximate}, provided that the sample numbers $N, N_{\mathsf{r}}\rightarrow\infty$. 
However, from the practical point of view, we can only have a finite number of samples. 
Thus, evaluating the viability of the approximate solution derived from finite samples becomes crucial. 
In the following, we discuss the theoretical result for the resampling-based approximation with finite $N$ and $N_{\mathsf{r}}$. 

For guaranteeing the feasibility of the approximate solution to the original problem \ref{eq:prob_orig} with a finite dataset, Hoeffding's inequality-based technique \cite{Luedtke, Pena, Shen2023_2} and statistical learning-based technique \cite{Campi:2008, Campi:2011, Garatti:2022} are commonly employed. However, the above techniques cannot be directly applied to cases involving contextual uncertainties. The aforementioned techniques require samples to be drawn from the exact probability distribution, which is unavailable in this setting.

Our approach adopts Hoeffding's inequality-based framework. 
First, we derive Hoeffding's inequality for the error between $\mathbb{P}_k(\bm{\theta}_k, \rho_{k-1})$ and its empirical counterpart $\widehat{\mathbb{P}}_k(\bm{\theta}_k, \rho_{k-1}, \mathcal{D}_N)$, with a finite sample size $N$. We then prove the probabilistic feasibility of the solution with finite $N$ and $N_{\mathsf{r}}$.

Define
\begin{equation}
\label{eq:def_barM}   
\overline{M}:=\inf_{\bm{\theta}_k\in\Theta_{\mathsf{c}},\rho_{k-1}\in\mathscr{P}_{k-1}^{\mathsf{c}}}\mathsf{d}\widehat{\mathbb{P}}_k(\bm{\theta}_k,\rho_{k-1},\mathcal{D}_N),
\end{equation}
where $$\mathsf{d}\widehat{\mathbb{P}}_k(\bm{\theta}_k,\rho_{k-1},\mathcal{D}_N):=\mathbb{P}_k(\bm{\theta}_k,\rho_{k-1})-\widehat{\mathbb{P}}_k(\bm{\theta}_k,\rho_{k-1},\mathcal{D}_N).$$ 
We give the following lemma about consistency for $\overline{M}$.

\begin{mylemma}
\label{lemma:pointwise_conver_M}
    \label{theo:uniform_convergence_M_compact}
    There exists two positive constants $\overline{A}_1$ and $\overline{A}_2$ such that 
    \begin{align}        
        \mathsf{Pr}&\left\{\overline{M} <-\kappa \right\}\leq \overline{A}_1\exp\left\{-\frac{\kappa^2}{\overline{A}_2 N^{\frac{2+\gamma}{4}}}\right\}, \label{eq:uniform_convergence_M}
    \end{align}
    for any $\kappa>0$.
\end{mylemma}

\begin{pf}
    The proof can be found in Appendix \ref{proof:theo_uniform_convergence_M_compact}. \qed
\end{pf}

Let $\kappa:=\alpha-\alpha_{\mathsf{s}}$ be the error of confidence level with a little abusement of notation, where $\alpha_{\mathsf{s}}$ is used in the constraint \eqref{eq:constraint_approximate} of the approximate problem \ref{eq:prob_approximate} and $\alpha$ is the desired confidence level in the original problem \ref{eq:prob_orig}. 
To ensure the probabilistic feasibility, $\alpha_{\mathsf{s}}$ should be set smaller than $\alpha,$ which makes $\kappa>0.$
Based on Lemma \ref{theo:uniform_convergence_M_compact}, the probabilistic feasibility with finite samples is summarized in the following theorem.

\begin{mytheo}
    \label{theo:finite_samples_feasibility}
    Suppose that Assumption \ref{assump:Pbb_prob_measure} holds. 
    For any $\rho_{k-1}\in\mathscr{P}_{k-1}^{\mathsf{c}}$, denote $\bm{\theta}^{\mathsf{v}}_k(\rho_{k-1})$ such that $\bm{\theta}^{\mathsf{v}}_k(\rho_{k-1})\notin \Theta_{k,\alpha}(\rho_{k-1})$. Then, 
    \begin{align}
        &\sup_{\rho_{k-1}\in\mathscr{P}_{k-1}^{\mathsf{c}}} \mathsf{Pr}
        \left\{\bm{\theta}^{\mathsf{v}}_k(\rho_{k-1})\in\widetilde{\Theta}_{k,\alpha_{\mathsf{s}}}(\rho_{k-1},\mathcal{D}_N,\widetilde{\mathcal{X}}^{N_{\mathsf{r}}}_k) \right\}\leq \nonumber \\
        &  \exp\left\{-2N_{\mathrm{r}}(\overline{M}+\kappa)^2\right\}+\overline{A}_1\exp\left\{-\frac{\kappa^2}{\overline{A}_2 N^{\frac{2+\gamma}{4}}}\right\}. \label{eq:finite_samples_feasibility}
    \end{align}
\end{mytheo}

\begin{pf}
    The proof can be found in Appendix \ref{proof:theo_finite_samples_feasibility}. \qed
\end{pf}

Theorem \ref{theo:finite_samples_feasibility} reveals that the proposed approximation with finite samples has a bounded probability of leading to an infeasible solution. The probability bound requires sufficient samples to ensure the accuracy of the LS-CDE and a smaller threshold $\alpha_{s}<\alpha$ to ensure constraint satisfaction in a high probability. Indeed, the sample number bound provided by Theorem \ref{theo:finite_samples_feasibility} increases as $1/\overline{M}^2,$ which is more conservative than scenario approach \cite{Garatti:2022}. 
However, the results of the scenario approach cannot be directly applied to this case due to the unavailability of samples in the exact distribution. 

%%%%%%%%%%%%%%%%%%%%%%%%%%%%%%%%%%%%%%%%%%%%%%%%%%%
%%%%%%%%%%%%%%%%%%%%%%%%%%%%%%%%%%%%%%%%%%%%%%%%%%%
\section{Example}\label{sec:numerical_example}

This section presents a numerical validation of the proposed method, alongside a comparison with two existing approaches. 
The selected case study involves computing the probabilistic reachable sets of a quadrotor system controlled by a model predictive controller (MPC). 
The results highlight the effectiveness of the proposed method and the superiority over the existing methods.

\subsection{Description}

Consider a nonlinear quadrotor dynamical system described by 
\begin{equation}
\label{eq:quadrotor_system}    \bm{\mathrm{x}}_{k+1}=\bm{\mathrm{A}}\bm{\mathrm{x}}_k+\bm{\mathrm{B}}(m_k)\bm{\mathrm{u}}_k+\bm{\mathrm{d}}(\bm{\mathrm{x}}_k,\varphi_k)+\bm{\omega}_k,
\end{equation}
with 
\begin{equation*}
    \bm{\mathrm{A}}=
    \begin{bmatrix}
    1 & \Delta t & 0 & 0 \\
    0 & 1 & 0 & 0 \\
    0 & 0 & 1 & \Delta t \\
    0 & 0 & 0 & 1
    \end{bmatrix},\ 
    \bm{\mathrm{B}}(m_k)=\frac{1}{m_k}
    \begin{bmatrix}
    \frac{\Delta t^2}{2} & 0 \\
    \Delta t & 0 \\
    0 & \frac{\Delta t^2}{2} \\
    0 & \Delta t
    \end{bmatrix},\ 
\end{equation*}
\begin{equation*}
    \bm{\mathrm{d}}(\bm{\mathrm{x}}_k,\varphi_k)=-\varphi_k
    \begin{bmatrix}
        \frac{\Delta t^2|v_{x,k}|v_{x,k}}{2}\\ 
        \Delta t|v_{x,k}|v_{x,k}\\ 
        \frac{\Delta t^2|v_{y,k}|v_{y,k}}{2}\\ 
        \Delta t|v_{y,k}|v_{y,k}
    \end{bmatrix},
\end{equation*}
where $\Delta t$ is the sampling time. The system state is represented as $\bm{\mathrm{x}}_k = [p_{x,k}, v_{x,k}, p_{y,k}, v_{y,k}]^\top$, where $[p_{x,k}, p_{y,k}]^{\top}$ indicates the position and $[v_{x,k}, v_{y,k}]^{\top}$ denotes the velocity components. 
The initial state of the system \eqref{eq:quadrotor_system} is given by $\bm{\mathrm{x}}_0 = [-0.5, 0, -0.5, 0]^\top$. 
The system input is defined by $\bm{\mathrm{u}}_k=[u_{x,k},u_{y,k}]^\top.$

The system \eqref{eq:quadrotor_system} includes a nonlinear component $\bm{\mathrm{d}}(\bm{\mathrm{x}}_k, \varphi_k)$. 
The mass of the quadrotor, $m_k$, and the drag coefficient, $\varphi_k$, are uncertain. 
Furthermore, the system is subject to an external disturbance, $\bm{\omega}_k$.
The uncertain variables are assumed to follow the Beta distributions: $(m_k - 0.75)/0.5 \sim \mathsf{Beta}(2,2)$ and $(\varphi_k - 0.4)/0.2 \sim \mathsf{Beta}(2,5)$, where $\mathsf{Beta}(a,b)$ represents the Beta distribution with shape parameters $a$ and $b$. 
The disturbance $\bm{\omega}_k$ is assumed to be a multivariate normal distribution $\mathcal{N}(\bm{0}^4,\Sigma)$. 
The diagonal elements of $\Sigma$ are specified as $\Sigma(1,1) = 0.01$, $\Sigma(2,2) = 0.75$, $\Sigma(3,3) = 0.01$, and $\Sigma(4,4) = 0.75$.
In this example, the contextual variable is defined by 
\begin{equation}
    \label{eq:context_uncertain_example}
    \bm{\mathrm{w}}_k=\bm{\mathrm{B}}(m_k)\bm{\mathrm{u}}_k+\bm{\mathrm{d}}(\bm{\mathrm{x}}_k, \varphi_k)+\bm{\omega}_k. 
\end{equation}

To clarify the role of the contextual variable ${\xi}_k$ in this example, we first describe the process for computing the system input $\bm{\mathrm{u}}_k$. 
In this example, an MPC is employed to navigate the quadrotor from its initial position to a desired goal point/region while ensuring the avoidance of unsafe regions. 
At each time $k$, the following MPC optimization is implemented:
\begin{align} 
    &\min_{\mathcal{A}_T} \,\, J(\mathcal{A}_T,\bm{\mathrm{z}}_0):=L_T(\bm{\mathrm{z}}_T)+\sum_{t=0}^{T-1}\ell(\bm{\mathrm{z}}_t,\bm{\mathrm{a}}_t),\label{eq:example_mpc}  \\
    &{\normalfont \mathsf{s.t.}}\quad  \bm{\mathrm{z}}_{t+1}=\bm{\mathrm{A}}\bm{\mathrm{z}}_t+\bm{\mathrm{B}}(\bar{m})\bm{\mathrm{u}}_t+\bm{\mathrm{d}}(\bm{\mathrm{z}}_t,\bar{\varphi}),\ \bm{\mathrm{z}}_0=\bm{\mathrm{x}}_k, \nonumber\\
    &\ \ \ \ \quad \land_{t=1}^{T-1}\bm{\mathrm{z}}_t\notin\mathcal{O},\ \bm{\mathrm{z}}_T\in\mathcal{Z}_{\mathsf{goal}},\ t=0,...,T-1,\nonumber
\end{align}
where $J(\mathcal{A}_T, \bm{\mathrm{z}}_0)$ is the cost function, which combines a terminal cost $L_T(\bm{\mathrm{z}}_T)$ and the cumulative stage costs $\ell(\bm{\mathrm{z}}_t, \bm{\mathrm{a}}_t)$ over the prediction horizon. The control sequence $\mathcal{A}_T$ is defined by $\mathcal{A}_T:=\{\bm{\mathrm{a}}_0,...,\bm{\mathrm{a}}_{T-1}\}.$ 
The unsafe region $\mathcal{O}$ is given by
\begin{align*}
    & \bm{\mathrm{z}}_1\leq 6.35,\ \bm{\mathrm{z}}_3\geq 3.35,\ \bm{\mathrm{z}}_1-0.2-\bm{\mathrm{z}}_3\geq 0, \\
    & \bm{\mathrm{z}}_1\geq 3.35,\ \bm{\mathrm{z}}_3\leq 6.35,\ \bm{\mathrm{z}}_1+0.2-\bm{\mathrm{z}}_3\leq 0.
\end{align*}

The destination set is $\mathcal{Z}_{\mathsf{goal}}=\{\bm{\mathrm{z}}|\|\bm{\mathrm{z}}-\bm{\mathrm{z}}_{\mathsf{goal}}\|\leq 2\},$ where $\bm{\mathrm{z}}_{\mathsf{goal}}=[10,\ 0,\ 10,\ 0]^\top$. 
The instantaneous cost $\ell(\cdot)$ is written by $\ell(\bm{\mathrm{z}}_t,\bm{\mathrm{a}}_t)=\|\bm{\mathrm{z}}_t-\bm{\mathrm{z}}_{\mathsf{goal}}\|^2_2+0.1\|\bm{\mathrm{a}}_t\|^2_2$ and the terminal cost is $L_T(\bm{\mathrm{z}}_T)=\|\bm{\mathrm{z}}_T-\bm{\mathrm{z}}_{\mathsf{goal}}\|^2_2$. 
The mass and coefficient take the maximum likelihood values, denoted by  $\bar{m}$ and $\bar{\varphi}$, respectively.

For given $\bm{\mathrm{z}}_0=\bm{\mathrm{x}}_k$, solving optimization problem \eqref{eq:example_mpc} yields optimal sequence $\mathcal{A}_{T}^*$. 
According to a receding horizon strategy, only the first input $\bm{\mathrm{a}}_0^*$ of $\mathcal{A}_{T}^*$ is applied to the system at time $k$. 
Note that the MPC forms a feedback control law $\bm{\mathrm{u}}_k =\bm{\mathrm{a}}^*_0$ based on the current state $\bm{\mathrm{x}}_k$\footnote{To avoid confusion, we do not use the term ``function" in this feedback control law because $\bm{\mathrm{a}}^*_0$ is not explicitly defined as a function of $\bm{\mathrm{x}}_k$. 
While $\bm{\mathrm{x}}_k$ determines the distribution of $\bm{\mathrm{a}}^*_0$, it does not uniquely determine the value of $\bm{\mathrm{a}}^*_0$ \cite{Shen_GMPC}.}. 
As discussed above, $\bm{\mathrm{w}}_k$ depends on $m_k$, $\varphi_k$, $\bm{\omega}_k$, and $\bm{\mathrm{x}}_k$, though it lacks a direct functional relationship with these variables due to the MPC control law $\bm{\mathrm{u}}_k =\bm{\mathrm{a}}^*_0$. 

In this example, the contextual variable is set as ${\xi}_k = \bm{\mathrm{x}}_k$, integrating the randomness of $m_k$, $\varphi_k$, and $\bm{\omega}_k$ directly into $\bm{\mathrm{w}}_k$\footnote{If the exact values of $m_k$, $\varphi_k$, and $\bm{\omega}_k$ are known at each time $k$, they could be explicitly included in the contextual variable ${\xi}_k$.}
. With this setup, we compare the performance of the following methods:
\begin{itemize}
    \item \textbf{Proposed}: The proposed method (Algorithm~\ref{alg:rsSAA});
    \item \textbf{RSA}: The resampling-based scenario method, based on the scenario approach \cite{Campi:2008, Shen2023_1} with samples that obtained by our resampling algorithm;
     \item \textbf{SCA}: The sample-based continuous approximation method proposed in~\cite{Shen2023_2}.
\end{itemize}

\subsection{Results and discussions}

\begin{figure*}[t]
\centering
\includegraphics[width = \hsize]{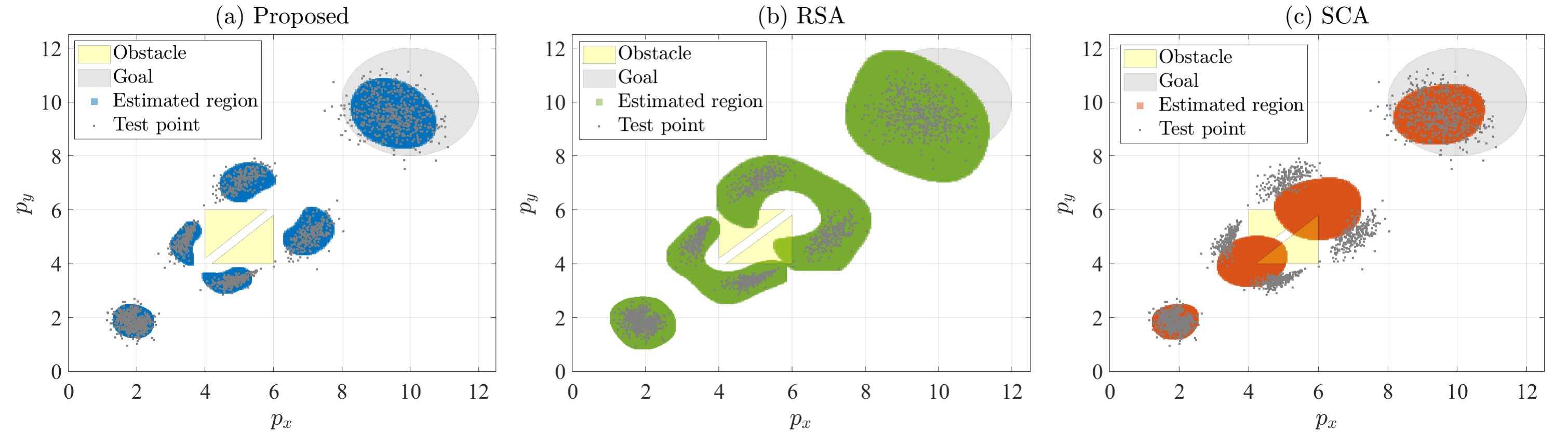}
\centering
\caption{The confidence regions obtained by \textbf{Proposed} with $N=1000,\ N_{\mathsf{r}}=500,\ d=2,\ \alpha_{\mathsf{s}}=0.235$. Here, the points are the data. The colored regions are the obtained confidence regions.}
\label{fig:example_three_methods}
\end{figure*}

\begin{figure}[t]
\centering
\includegraphics[width = \hsize]{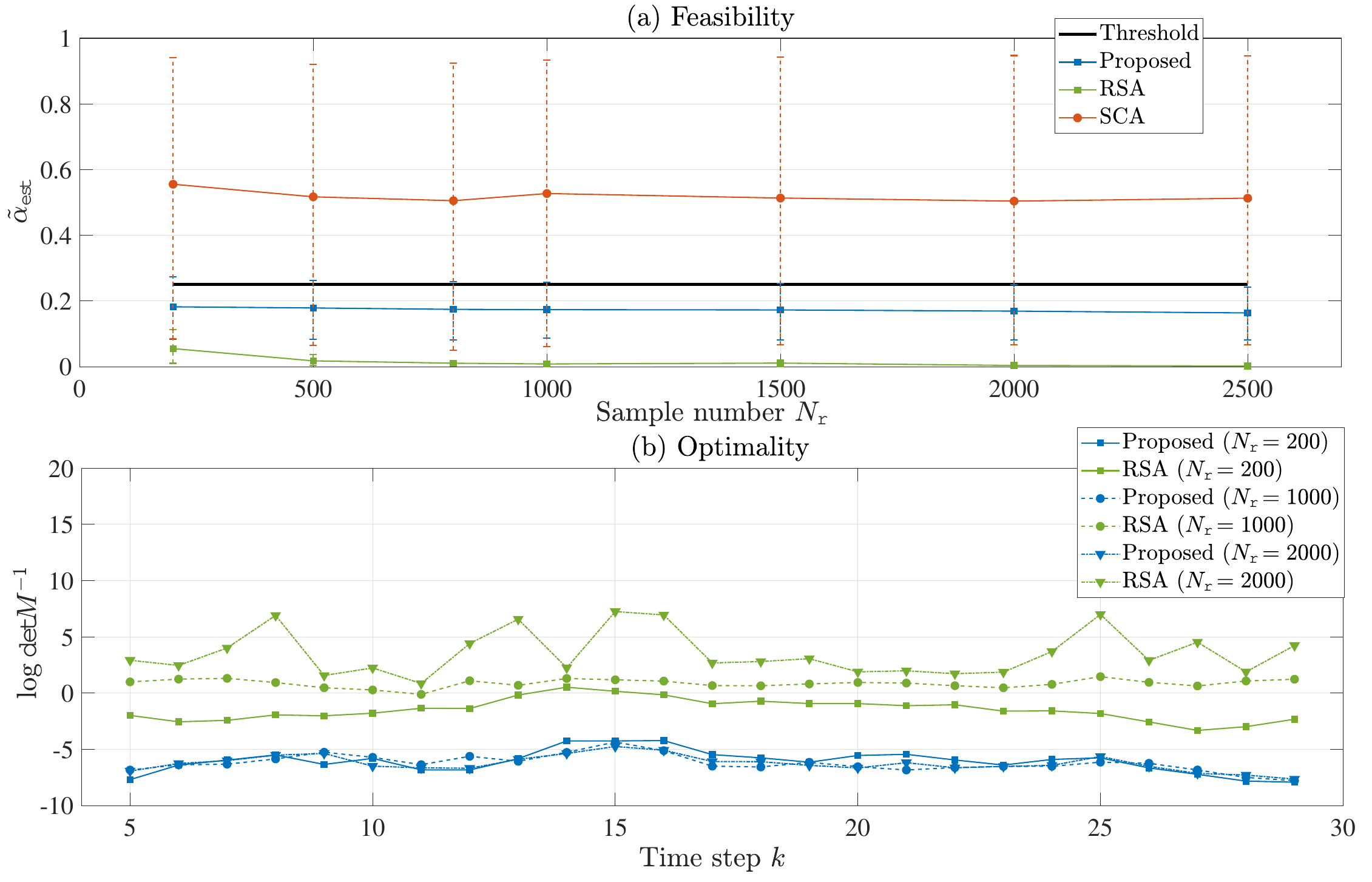}
\centering
\caption{The confidence regions obtained by \textbf{Proposed} with $N=1000,\ N_{\mathsf{r}}=500,\ d=2,\ \alpha_{\mathsf{s}}=0.235$. Here, the points are the data. The colored regions are the obtained confidence regions.}
\label{fig:statistics}
\end{figure}

In this example, our objective is to compute the minimal ($1-\alpha$)-reachable polynomial sublevel set of $\bm{\mathrm{x}}_k$ for $k=5, \dots, 29$, starting from the initial state $\bm{\mathrm{x}}_0$. 
The probability threshold is set to $\alpha = 0.25$. 
The training set $\mathcal{D}_N$ is generated by running the system \eqref{eq:context_uncertain_example} under MPC for 3,000 time steps. 
For each method, the number of resampled data points used for probabilistic constraint approximation is $N_{\mathsf{r}} \in \{200, 500, 800, 1000, 1500, 2000, 2500 \}$ and the degree is set as $d=2$. 
Both \textbf{Proposed} and \textbf{SCA} use a slightly lower probability threshold, $\alpha_{\mathsf{s}} = 0.235$, compared to $\alpha = 0.25$. 
For \textbf{RSA}, the resampled data points are identical to those in \textbf{Proposed}, with the key difference being that \textbf{RSA} requires the constraint to be satisfied for all resampled data points. 
For \textbf{SCA}, the dataset $\mathcal{D}_N^{\mathrm{W}}$ is constructed directly from the samples $\bm{\mathrm{w}}_{\mathsf{s}}^{(i)}, i=1, \dots, N$ in $\mathcal{D}_N$.

Fig. \ref{fig:example_three_methods} (a), (b), (c) illustrate the confidence regions at time $k=5,\ 10,\ 15,\ 29$ by \textbf{Proposed}, \textbf{RSA}, and \textbf{SCA}, respectively. 
As time $k$ increases, the confidence region determined by \textbf{SCA} starts to diverge from the observed data. 
This divergence, attributed to their disregard for the conditional probability density of $\bm{\mathrm{w}}_{k}$, becomes increasingly pronounced over time, particularly near the unsafe region. 
In contrast, \textbf{Proposed} and \textbf{RSA}, which account for the conditional probability density of $\bm{\mathrm{w}}_k$, maintain confidence regions that closely align with the observed data. 
Notably, \textbf{RSA} produces significantly larger confidence regions compared to \textbf{Proposed}. This is due to the conservativeness of scenario approach, which requires the confidence region to encompass all training data points, leading to an overestimation. 

An extensive statistical evaluation was performed using Monte-Carlo simulations across three methods. Each method underwent 1,000 runs. 
From these simulations, we accessed the estimated risk level $\tilde{\alpha}_{\mathsf{s}}$ computed from the test data points. 
An $\tilde{\alpha}_{\mathsf{s}}$ value close to $0.25$ indicates that the obtained confidence region closely matches the desired performance. 

Fig. \ref{fig:statistics} (a) and (b) show the comparison results of three methods on the feasibility and optimality, respectively.
As shown by the red lines in Figure \ref{fig:statistics} (a), \textbf{SCA} fails to achieve a conservative $\tilde{\alpha}_{\mathsf{s}}$, deviating significantly from the ideal value of $\alpha = 0.25$. 
The average value is well above $\alpha = 0.25$, with most of the $99\%$ confidence intervals also exceeding this threshold.
This is due to its disregard for the conditional probability density in data sampling, leading to sample-based constraints deviating from the no-biased approximations. 
Such approximations result in biased confidence regions, inadequately addressing risk even with a large confidence region.
In contrast, \textbf{RSA} (green lines in Fig. \ref{fig:statistics}) and \textbf{Proposed} (blue lines in Fig. \ref{fig:statistics}) satisfy the risk threshold with high confidence. 
In particular, the proposed resampling strategy ensures that the sample-based approximation aligns closely with the original problem as proved by Theorem~\ref{theo:almost_uniform_convergence}. 
As a result, in Fig. \ref{fig:statistics} (a), with \textbf{Proposed}, the estimated risk levels $\tilde{\alpha}_{\mathsf{s}}$ are consistently close to the desired value of $\alpha=0.25$.

For the optimality comparison, we focus on \textbf{Proposed} and \textbf{RSA}, as \textbf{SCA} fails to provide feasible solutions, making an optimality comparison with infeasible solutions irrelevant.
As shown in Fig. \ref{fig:statistics} (b), \textbf{RSA} yields significantly higher obtained optimal values than \textbf{Proposed}, which corresponds to larger confidence regions in Fig. \ref{fig:example_three_methods} (a) and (b). 
Conversely, \textbf{Proposed} effectively mitigates this bias through the implementation of the LS-CDE and a tailored resampling strategy. 

\section{Conclusions}
\label{sec:conclusions}

This paper proposes a novel method for computing probabilistic reachable sets of future state trajectories in a stochastic nonlinear system with contextual uncertainties. 
These reachable sets are characterized as minimum-volume polynomial sublevel sets with chance constraints. 
The contextual uncertainties follow a probability density conditioned on the current state, and the probabilistic reachable sets are derived by solving a stochastic optimization problem. 
Existing sample-based approximation methods cannot directly solve this problem, as they do not account for conditional probability densities.

We have proposed a resampling-based approximation of the original problem to address this limitation. 
First, we implement least-squares conditional density estimation to learn the conditional probability density. 
Based on the estimated density, we then design a resampling strategy to generate new samples to approximate the original problem. 
Theoretically, we prove almost uniform convergence of the sample-based approximation, demonstrating that the proposed method yields an optimal solution consistent with the original problem, provided sufficient data samples are available. 
A numerical example benchmarks the proposed method against two existing approaches, showcasing its ability to correct the bias found in sample-based methods that overlook conditional probability densities. 
As future work, the probabilistic reachable sets obtained in this study will be applied to developing stochastic model predictive control for safety-critical systems.

\begin{ack}                               % Place acknowledgements
    Dr Xun Shen acknowledges support from Japan Society for the Promotion of Science via JSPS Kakenhi (24K16752), Research Organization of Information and Systems via Strategic Research Project (2023-SRP-06), and Osaka University Institute for Datability Science (IDS Interdisciplinary Collaboration Project). Dr Ye Wang acknowledges support from the Australian Research Council through the Discovery Early Career Researcher Award (DE220100609).  % here.
\end{ack}

%\bibliographystyle{plain}        % Include this if you use bibtex 
%\bibliography{autosam}           % and a bib file to produce the 

\begin{thebibliography}{99}

\bibitem{Haesaert}
S. Haesaert, P.M. Van denHof, and A. Abate, “Data-driven and model-based verification via Bayesian identification and reachability analysis.” \textit{Automatica}, vol. 79, pp. 115–126, 2017.

\bibitem{Shen2023_1}
X. Shen, T. Ouyang, Y. Zhang, X. Zhang, “Computing probabilistic bounds on state trajectories for uncertain systems.” \textit{IEEE Transactions on Emerging Topics in Computational Intelligence }, vol. 7, no. 1, pp. 285 - 290, 2023.

\bibitem{Shen2023_2}
X. Shen, “Sample-based continuous approximation for computing probabilistic boundary of future state trajectory in uncertain dynamical systems.” \textit{IEEE Transactions on Emerging Topics in Computational Intelligence }, vol. 8, no. 2, pp. 1110-1117, 2024.

\bibitem{Mirko:2021}
F. Mirko and T. Alamo. "Probabilistic reachable and invariant sets for linear systems with correlated disturbance." \textit{Automatica}, vol. 132, 109808, 2021.

\bibitem{Bloemers}
T. Bloemers, T. Oomen, and R. Toth, “Frequency response data-driven LPV controller synthesis for mimo systems.” \textit{IEEE Control Systems Letters}, vol. 6, pp. 2264–2269, 2022.

\bibitem{Karimi}
A. Karimi and C. Kammer, “A data-driven approach to robust control of multivariable systems by convex optimization.” \textit{Automatica}, vol. 85, pp. 227–233, 2017.

\bibitem{Zanon}
M. Zanon and S. Gros, “Safe reinforcement learning using robust MPC.” \textit{IEEE Transactions on Automatic Control}, vol. 66, no. 8, pp. 3638–3652, 2021.

\bibitem{Akella:2022}
P. Akella and A. Ames, “A barrier-based scenario approach to verifying safety-critical systems.” \textit{IEEE Robotics and Automation Letters}, vol. 7, no. 4, pp. 11062 – 11069, 2022.

\bibitem{A:Wan}
Y. Wan, T. Keviczky, M. Verhaegen, and F. Gustafsson, “Data-driven robust receding horizon fault estimation.” \textit{Automatica}, vol. 71, pp. 210–221, 2016.

\bibitem{Dabbene:2015}
F. Dabbene, D. Henrion, C. Lagoa, and P. Shcherbakov, “Randomized approximations of the image set of nonlinear mappings with applications to filtering.” \textit{IFAC-PapersOnLine}, vol. 48, no. 14, pp. 37–42,
2015.

\bibitem{Calafiore:2006}
G. Calafiore and M. C. Campi, “The scenario approach to robust control design.” \textit{IEEE Transactions on Automatic Control}, vol. 51, no. 5, pp. 742–753, 2006.

\bibitem{Shen2023_TNNLS}
X. Shen, T. Ouyang, N. Yang, J. Zhuang, “Sample-based neural approximation approach for probabilistic constrained programs,” \textit{IEEE Transactions on Neural Networks and Learning Systems}, vol. 34, no. 2, pp. 1058-1065, 2023.

\bibitem{CampiInterval}
M. Campi, G. Calafiore, and S. Garatti, “Interval predictor models:
identification and reliability.” \textit{Automatica}, vol. 45, no. 2, pp. 382–392, 2009.

\bibitem{Margellos}
K. Margellos, P. Goulart, and J. Lygeros, “On the road between robust optimization and the scenario approach for chance constrained optimization problems.” \textit{IEEE Transactions on Automatic Control}, vol. 59, no. 8, pp. 2258–2263, 2014.

\bibitem{Au:Alamo}
T. Alamo, J. M. Bravo, M. J. Redondo, and E. F. Camacho, “A set-membership state estimation algorithm based on DC programming.” \textit{Automatica}, vol. 44, pp. 216–224, 2008.

\bibitem{OE:Kishida}
M. Kishida and R. B. Braatz, “Ellipsoidal bounds on state trajectories for discrete-time systems with linear fractional uncertainties.” \textit{Optim. Eng.}, vol. 16, pp. 695–711, 2015.

\bibitem{Campi:2011}
M. Campi and S. Garatti, “A sampling-and-discarding approach to chance-constrained optimization: feasibility and optimality.” \textit{Journal of Optimization Theory and Applications}, vol. 148, no. 2, pp. 257–280, 2011.

\bibitem{Campi:2018}
M. Campi, S. Garatti, and F.A. Ramponi “A general scenario theory for nonconvex optimization and decision making.” \textit{IEEE Transactions on Automatic Control}, vol. 63, no. 12, pp. 4067-4078, 2018.

\bibitem{Garatti:2022}
S. Garatti and M. Campi, “Risk and complexity in scenario optimization.” \textit{Mathematical Programming}, vol. 191, pp. 243–279, 2022.

\bibitem{Romao2022}
L. Romao, A. Papachristodoulou, and K. Margellos, “On the exact feasibility of convex scenario programs with discarded constraints.” \textit{IEEE Transactions on Automatic Control}, vol. 68, no. 4, pp. 1986–2001, 2022.

\bibitem{Wang:Auto2024}
X. Wang and E. Weyer, “Scenario based optimisation over uncertain system identification models.” \textit{Automatica}, vol. 165, no. 2, pp. 243–279, 2024.

\bibitem{Luedtke}
J. Luedtke and S. Ahmed, “A sample approximation approach for optimization with probabilistic constraints.” \textit{SIAM Journal on Optimization}, vol. 19, no. 2, pp. 674-699, 2008.

\bibitem{Shen2023_JOTA}
X. Shen and S. Ito, “Approximate methods for solving chance-constrained linear programs in probability measure space.” \textit{Journal of Optimization Theory and Applications}, vol. 200, pp. 150-177, 2024.

\bibitem{Dabbene:2017}
F. Dabbene, D. Henrion, and C. Lagoa, “Simple approximations of semialgebraic sets and their applications to control.” \textit{Automatica}, vol. 78, pp. 110-118, 2017.

\bibitem{Lasserre:2015}
J.B. Lasserre, “Level sets and non Gaussian integrals of positively homogeneous functions.” \textit{International Game Theory Review}, vol. 15, no. 1.1540001, 2015.

\bibitem{Morozov}
A.Y. Morozov and S.R. Shakirov, “New and old results in resultant theory.” \textit{Theoretical and Mathematical Physics}, vol. 163, no. 2, pp. 587-617, 2010.

\bibitem{Magnani}
A. Magnani, S. Lall, and S. Boyd, “Tractable fitting with convex polynomials via sum-of-squares,” in \textit{44th IEEE Conference on Decision and Control, and the European Control Conference}, 2005.

\bibitem{Chen1995}
C. Chen and O. Mangasarian, “Smoothing methods for convex inequalities and linear complementarity problems.” \textit{Mathematical Programming}, vol. 71, no. 1, pp. 51–69, 1995.

\bibitem{Geletu:2017}
A. Geletu, A. Hoffmann, M. Kloppel, and P. Li, “An inner-outer approximation approach to chance constrained optimization.” \textit{SIAM Journal on Optimization}, vol. 27, no. 3, pp. 1834–1857, 2017.

\bibitem{Soloperto}
R. Soloperto, M.A. Muller, S. Trimpe, and F. Allgower, ``Learning-based robust model predictive control with state-dependent uncertainty", \textit{IFAC PapersOnLine}, vol. 51, no. 20, pp. 442-447, 2018.

\bibitem{Menner}
M. Menner and K. Berntorp, ``Gaussian processes with state-dependent noise for stochastic control." \textit{2023 IEEE Conference on Control Technology and Applications (CCTA)}, Bridgetown, Barbados, 16-18 August 2023. 

\bibitem{Zhao_Pan}
P. Zhao, I. Kolmanovsky, and N. Hovakimyan, “Integrated adaptive control and reference governors for constrained systems with state-dependent uncertainties.” \textit{IEEE Transactions on Automatic Control}, vol. 69, no. 5, pp. 3158-3173, 2024.

\bibitem{Kohler}
J. Kohler, R. Soloperto, M.A. Muller, and F. Allgower, “A computationally efficient robust model predictive control framework for uncertain nonlinear systems.” \textit{IEEE Transactions on Automatic Control}, vol. 66, no. 2, pp. 794-801, 2021.

\bibitem{Arrigo}
A. Arrigo et al., “Wasserstein distributionally robust chance-constrained optimization for energy and reserve dispatch: An exact and physically-bounded formulation.” \textit{European Journal of Operational Research}, vol. 296, no. 1, pp. 304-322, 2022.

\bibitem{Biau}
G. Biau, “Analysis of a random forests model,” \textit{Journal of Machine Learning Research}, vol. 13, pp. 1063--1095, 2012.

\bibitem{Doring}
M. Doring, L. Gyorfi, and H. Walk, “Rate of convergence of k-nearest-neighbor classification rule,” \textit{Journal of Machine Learning Research}, vo. 18, pp. 8485--8500, 2017.

\bibitem{Capinski}  
M. Capinski and E. Kopp, \textit{Measure, Integral, and Probability}, Springer-Verlag, London, 2004.

\bibitem{Hernandez}  
 O. Hernandez-Lerma and J. Lasserre, {\it Discrete-Time Markov Control Processes: Basic Optimality Criteria}, Springer, 1996.

\bibitem{Shen_GMPC}
X. Shen, “Generative model predictive control: approximating {MPC} law with generative models.” \textit{IEEE Transactions on Emerging Topics in Computational Intelligence}, DOI: 10.1109/TETCI.2024.3358096.

\bibitem{Kozhasov}
K. Kozhasov and J.B. Lasserre, “Nonnegative forms with sublevel sets of minimal volume.” \textit{Mathematical Programming}, vol. 193, pp. 485-498, 2022.

\bibitem{Rahimian}
H. Rahimian and B. Pagnoncelli, “Data-driven approximation of contextual chance-constrained stochastic programs.” \textit{SIAM Journal on Optimization}, vol. 33, no. 3, pp. 2248–2274, 2023.

\bibitem{Pena}
A. Pena-{Ordieres}, J.R. Luedteke, and A. Wachter, “Solving chance-constrained problems via a smooth sample-based nonlinear approximation.” \textit{SIAM Journal on Optimization}, vol. 30, no. 3, pp. 2221-2250, 2020.

\bibitem{Kibzun}
A. Kibzun and Y. Kan, “Stochastic Programming Problems with Probability and Quantile Functions.” \textit{Journal of the Operational Research Society}, vol. 48, pp. 846-856, 1997.

\bibitem{Wolff}
R.C.L. Wolff, Q. Yao, and P. Hall, “Methods for estimating a conditional distribution function.” \textit{Journal of the American Statistical Association }, vol. 94, no. 445, pp. 154 - 163, 1999.

\bibitem{Sugiyama_AISM}
M. Sugiyama, T. Suzuki, S. Nakajima, H. Kashima, P. vonBunau, and M. Kawanabe, “Direct importance estimation for covariate shift adaptation.” \textit{Annals of the Institute of Statistical Mathematics }, vol. 60, pp. 699 - 746, 2008.

\bibitem{Kanamori}
T. Kanamori, S. Hido, and M. Sugiyama, “A least-squares approach to direct importance estimation.” \textit{Journal of Machine Learning Research}, vol. 10, pp. 1391-1445, 2009.

\bibitem{Sugiyama}
M. Sugiyama, I. Takeuchi, t. Suzuki, T. Kanamori, H. Hachiya, and D. Okanohara, “Least-squares conditional density estimation.” \textit{IEICE Transactions on Information and Systems }, vol. E93-D, no. 3, pp. 583 - 594, 2010.

\bibitem{Campi:2008}
M. Campi and S. Garatti, “The exact feasibility of randomized solutions of uncertain convex programs.” \textit{SIAM Journal on Optimization}, vol. 19, no. 3, pp. 1211–1230, 2008.

\bibitem{Carlin}
B.P. Carlin, N.G. Polson, and D.S. Stoffer, “A Monte Carlo approach to nonnormal and nonlinear state-space modeling.” \textit{Journal of the American Statistical Association }, vol. 87, no. 418, pp. 493-500, 1992. 

\bibitem{Wheeden}
R. Wheeden, A. Zygmund, \textit{Measure and Integral: An Introduction to Real Analysis}, CRC Press, Boca Raton, 1977.

\bibitem{Williams}
D. Williams, \textit{Probability with Martingales}, Cambridge University Press, New York, 1991.

\bibitem{Bertsekas_Probability}
D.P. Bertsekas, J.N. Tsitsiklis, \textit{Introduction to Probability, Second Edition}, Athena Scientific, Nashua, 2008.

\bibitem{Billingsley}  
P. Billingsley, \textit{Convergence of probability measures}, John Wiley $\&$ Sons, 1999.

\bibitem{Hoeffding}
W. Hoeffding, “Probability inequalities for sums of bounded random variables.” \textit{Journal of the American Statistical Association}, vol. 58, pp. 13–30, 1963.

\end{thebibliography}
                                 % bibliography (preferred). The
                                 % correct style is generated by
                                 % Elsevier at the time of printing.

%\begin{thebibliography}{99}     % Otherwise use the  
                                 % thebibliography environment.
                                 % Insert the full references here.
                                 % See a recent issue of Automatica 
                                 % for the style.
%  \bibitem[Heritage, 1992]{Heritage:92}
%     (1992) {\it The American Heritage. 
%     Dictionary of the American Language.}
%     Houghton Mifflin Company.
%  \bibitem[Able, 1956]{Abl:56}
%     B.~C.~Able (1956). Nucleic acid content of macroscope. 
%     {\it Nature 2}, 7--9. 
%  \bibitem[Able {\em et al.}, 1954]{AbTaRu:54}   
%     B.~C. Able, R.~A. Tagg, and M.~Rush (1954).
%     Enzyme-catalyzed cellular transanimations.
%     In A.~F.~Round, editor, 
%     {\it Advances in Enzymology Vol. 2} (125--247). 
%     New York, Academic Press.
%  \bibitem[R.~Keohane, 1958]{Keo:58}
%     R.~Keohane (1958).
%     {\it Power and Interdependence: 
%     World Politics in Transition.}
%     Boston, Little, Brown \& Co.
%  \bibitem[Powers, 1985]{Pow:85}
%     T.~Powers (1985).
%     Is there a way out?
%     {\it Harpers, June 1985}, 35--47.

%\end{thebibliography}

\bibliographystyle{plain}

\appendix

\section{Proof of Theorem \ref{theo:convergence_tild_bbP_k}}
\label{proof:theo_convergence_tild_bbP_k}

In this proof, to simplify notation, $\bm{\theta}_k$ is shortly written as $\bm{\theta}$. 
Almost uniform convergence of $ \widetilde{\mathbb{P}}_k(\bm{\theta},\rho_{k-1},\mathcal{D}_N,\widetilde{\mathcal{X}}_k^{N_\mathsf{r}})$ to $\mathbb{P}_k(\bm{\theta},\rho_{k-1})$ can be obtained by showing
\begin{itemize}
    \item[(a)] Almost uniform convergence of $ \widetilde{\mathbb{P}}_k(\bm{\theta},\rho_{k-1},\mathcal{D}_N,\widetilde{\mathcal{X}}_k^{N_\mathsf{r}})$ to $\widehat{\mathbb{P}}_k(\bm{\theta},\rho_{k-1},\mathcal{D}_N)$ over $(\bm{\theta},\rho_{k-1})\in\Theta_{\mathsf{c}}\times\mathscr{P}_{k-1}^{\mathsf{c}}$;
    \item[(b)] Almost uniform convergence of $\widehat{\mathbb{P}}_k(\bm{\theta},\rho_{k-1},\mathcal{D}_N)$ to $\mathbb{P}_k(\bm{\theta},\rho_{k-1})$ over $(\bm{\theta},\rho_{k-1})\in\Theta_{\mathsf{c}}\times\mathscr{P}_{k-1}^{\mathsf{c}}$. 
\end{itemize}

We first prove the statement (a). Algorithm \ref{alg:rsSAA} essentially extracts sample set $\widetilde{\mathcal{X}}_k^{N_{\mathsf{r}}}$ by Gibbs sampling based on $\hat{p}^{\mathsf{c}}_{\bm{\mathrm{W}}_j}(\bm{\mathrm{w}}_j|\xi_j),\ j=0,...,k-1$, which is identical as $\hat{p}^{\mathsf{c}}_{\bm{\mathrm{W}}}(\bm{\mathrm{w}}|\xi_j)$ by replacing $\xi$ by $\xi_j$ in \eqref{eq:LS_CDE}. 
During the iterative computation, in $\xi_j=\{\bm{\mathrm{x}}_j,\bm{\mathrm{v}}_j\}$, where $\bm{\mathrm{v}}_j$ is directly obtained from $\rho_{k-1}.$    
Therefore, we can regard the Gibbs sampling-based sample set $\widetilde{\mathcal{X}}_k^{N_{\mathsf{r}}}$ is extracted obeying a conditional probability distribution $\hat{p}^{\mathsf{c}}_{\bm{\mathrm{X}}_k}(\bm{\mathrm{x}}_k|\rho_{k-1})$ \cite{Carlin}. 
By applying \cite[Theorem 3.4]{Pena}, we have that, for any $\rho_{k-1}\in\mathscr{P}_{k-1}$, as $N_{\mathsf{r}}\rightarrow\infty$, with probability 1,
\begin{equation}
    \label{eq:convergence_tild_bbP_k_k_p1_pointwise}
    \sup_{\bm{\theta}\in\Theta_{\mathsf{c}}}\left|\widetilde{\mathbb{P}}_k-\widehat{\mathbb{P}}_k\right|\rightarrow 0.
\end{equation}
Here, $\widetilde{\mathbb{P}}_k$ and $\widehat{\mathbb{P}}_k$ are abbreviations for $\widetilde{\mathbb{P}}_k(\bm{\theta},\rho_{k-1},\mathcal{D}_N,\widetilde{\mathcal{X}}_k^{N_\mathsf{r}})$ and $\widehat{\mathbb{P}}_k(\bm{\theta},\rho_{k-1},\mathcal{D}_N)$.
Note that \eqref{eq:convergence_tild_bbP_k_k_p1_pointwise} shows the uniform convergence on $\Theta_{\mathsf{c}}$ for any $\rho_{k-1}\in\mathscr{P}_{k-1}$, namely, the pointwise convergence on $\mathscr{P}_{k-1}$ is given. 
Since $\widetilde{\mathbb{P}}_k(\bm{\theta},\rho_{k-1},\mathcal{D}_N,\widetilde{\mathcal{X}}_k^{N_\mathsf{r}})$ is a measurable function on $\mathscr{P}_{k-1}$, by the Egorov's theorem \cite[page 57]{Wheeden}, pointwise convergence leads to almost uniform convergence, and we have that, as $N_{\mathsf{r}}\rightarrow\infty$, with probability 1,
\begin{equation}
        \label{eq:convergence_tild_bbP_k_k_p1}
        \sup_{\bm{\theta}\in\Theta_{\mathsf{c}},\rho_{k-1}\in\mathscr{P}_{k-1}^{\mathsf{c}}}\left|\widetilde{\mathbb{P}}_k-\widehat{\mathbb{P}}_k\right|\rightarrow 0,
\end{equation}
Here, $\widetilde{\mathbb{P}}_k$ and $\widehat{\mathbb{P}}_k$ are abbreviations for $\widetilde{\mathbb{P}}_k(\bm{\theta},\rho_{k-1},\mathcal{D}_N,\widetilde{\mathcal{X}}_k^{N_\mathsf{r}})$ and $\widehat{\mathbb{P}}_k(\bm{\theta},\rho_{k-1},\mathcal{D}_N)$.
Besides, $\mathscr{P}_{k-1}^{\mathsf{c}}$ can be a compact subset of $\mathscr{P}_{k-1}$ such that $\mu_{\rho}(\mathscr{P}_{k-1}\setminus\mathscr{P}_{k-1}^{\mathsf{c}})<\epsilon$ for any $\epsilon>0,$ where $\mu_{\rho}$ is any probability measure on $\mathscr{P}_{k-1}$ associated with positive probability density everywhere.

Then, we prove the statement (b) and start with $k=1$. Lemma~\ref{lemma:convergence_LS_CDE} implies the pointwise convergence of LS-CDF $\hat{p}^{\mathsf{c}}_{\bm{\mathrm{W}}}(\cdot|\xi)$ to $p^{\mathsf{c}}_{\bm{\mathrm{W}}}(\cdot|\xi)$ for every $\xi\in\bm{\Xi}.$ 
By applying the Scheffe's lemma \cite[page 55]{Williams}, we can obtain that the probability measure associated with $\hat{p}^{\mathsf{c}}_{\bm{\mathrm{W}}}(\cdot|\xi)$ weakly converges to the one associated with $p^{\mathsf{c}}_{\bm{\mathrm{W}}}(\cdot|\xi)$. 
Let 
\begin{equation*}
    \overline{\mathcal{W}}_{\bm{\mathrm{x}}_0}^{\bm{\theta}}:=\left\{\bm{\mathrm{w}}_0\in\mathcal{W}:q\left(f(\bm{\mathrm{x}}_0,\bm{\mathrm{w}}_0),\bm{\theta}\right)\leq 1\right\}.
\end{equation*}
be a set of $\bm{\mathrm{w}}_0$ determined by given $\bm{\mathrm{x}}_0$ and $\bm{\theta}.$ 
Notice that $\mathbb{P}_1(\bm{\theta},\rho_0)$ can be equivalently written by
\begin{align*}
    \mathbb{P}_1(\bm{\theta},\rho_0)&=\int_{\mathcal{X}} \mathbb{I}_{1}(q(\bm{\mathrm{x}}_1,\bm{\theta})) p^{\mathsf{c}}_{\bm{\mathrm{X}}_1}(\bm{\mathrm{x}}_1|\xi_0) \mathsf{d}\bm{\mathrm{x}}_1 \\
    &=\int_{\mathcal{W}} \mathbb{I}_{1}(q(f(\bm{\mathrm{x}}_0,\bm{\mathrm{w}}_0),\bm{\theta})) p^{\mathsf{c}}_{\bm{\mathrm{W}}_0}(\bm{\mathrm{w}}_0|\xi_0) \mathsf{d}\bm{\mathrm{w}}_0 \\
    &=\mathsf{Pr}_{\bm{\mathrm{W}}_0\sim p^{\mathsf{c}}_{\bm{\mathrm{W}}_0}(\cdot|\xi_0)}\left\{\bm{\mathrm{W}}_0\in\overline{\mathcal{W}}_{\bm{\mathrm{x}}_0}^{\bm{\theta}}\right\}.
\end{align*}

Note that $\rho_0$ equals $\xi_0$. In the same way, we can rewrite $\widehat{\mathbb{P}}_1(\bm{\theta},\rho_0,\mathcal{D}_N)$ as
\begin{equation*}
    \widehat{\mathbb{P}}_1(\bm{\theta},\rho_0,\mathcal{D}_N)=\mathsf{Pr}_{\bm{\mathrm{w}}_0\sim \hat{p}^{\mathsf{c}}_{\bm{\mathrm{W}}_0}(\cdot|\xi_0)}\left\{\bm{\mathrm{W}}_0\in\overline{\mathcal{W}}_{\bm{\mathrm{x}}_0}^{\bm{\theta}}\right\}.
\end{equation*}

By Assumption \ref{assump:Pbb_prob_measure}, $\overline{\mathcal{W}}_{\bm{\mathrm{x}}_0}^{\bm{\theta}}$ is a continuity set of $\bm{\mathrm{W}}_0\sim p^{\mathsf{c}}_{\bm{\mathrm{W}}_0}(\cdot|\xi_0)$ and $\mathbb{P}_1(\bm{\theta},\rho_0)$ is continuous function of $\bm{\theta}$. 
We choose the basis function $\bm{\phi}(\cdot)$ to make $\hat{p}^{\mathsf{c}}_{\bm{\mathrm{X}}_k}(\bm{\mathrm{x}}_k|\xi_0)$ satisfying Assumption \ref{assump:Pbb_prob_measure}, for example, choosing Gaussian function. 
Thus, $\overline{\mathcal{W}}_{\bm{\mathrm{x}}_0}^{\bm{\theta}}$ is also a continuity set of $\bm{\mathrm{W}}_0\sim \hat{p}^{\mathsf{c}}_{\bm{\mathrm{W}}_0}(\cdot|\xi_0)$ and $\widehat{\mathbb{P}}_1(\bm{\theta},\rho_0,\mathcal{D}_N)$ is continuous function of $\bm{\theta}$. 
By the Portmanteau Theorem (Theorem 2.1 of \cite{Billingsley}), we have $\widehat{\mathbb{P}}_1(\bm{\theta},\rho_0,\mathcal{D}_N)\rightarrow\mathbb{P}_1(\bm{\theta},\rho_0)$ for every $\bm{\theta}\in\Theta$ and $\rho_0\in\mathscr{P}_0.$ Due to the continuity of $\mathbb{P}_1(\bm{\theta},\rho_0)$ and $\widehat{\mathbb{P}}_1(\bm{\theta},\rho_0,\mathcal{D}_N)$ on $\Theta$, for every $\rho_0,$ we have that, as $N\rightarrow\infty$, with probability 1, 
\begin{equation}
    \label{eq:convergence_tild_bbP_k_1_p2_pointwise}    \sup_{\bm{\theta}\in\Theta_{\mathsf{c}}}\left|\widehat{\mathbb{P}}_1(\bm{\theta},\rho_0,\mathcal{D}_N)-\mathbb{P}_1(\bm{\theta},\rho_0)\right|\rightarrow 0.
\end{equation}

By also applying Egorov's theorem \cite[page 57]{Wheeden}, the pointwise convergence on $\mathscr{P}_{0}$ leads to almost uniform convergence, and we have that, as $N\rightarrow\infty$, with probability~1,
\begin{equation}
        \label{eq:convergence_tild_bbP_k_1_p2}
        \sup_{\bm{\theta}\in\Theta_{\mathsf{c}},\rho_0\in\mathscr{P}_{0}^{\mathsf{c}}}\left|\widehat{\mathbb{P}}_1(\bm{\theta},\rho_0,\mathcal{D}_N)-\mathbb{P}_1(\bm{\theta},\rho_0)\right|\rightarrow 0.
\end{equation}

Here, $\mathscr{P}_{0}^{\mathsf{c}}$ can be a compact subset of $\mathscr{P}_{0}$ such that $\mu_{\rho_0}(\mathscr{P}_{0}\setminus\mathscr{P}_{0}^{\mathsf{c}})<\epsilon$ for any $\epsilon>0,$ where $\mu_0$ is an arbitrary measure on $\mathscr{P}_0.$ 

For $k\geq 1,$  let us define
\begin{align*}
    \mathbb{P}^+_{k}(\bm{\theta},\xi_k):=& \int_{\mathcal{W}} \Scale[0.96]{\mathbb{I}_{1}(q(f(\bm{\mathrm{x}}_k,\bm{\mathrm{w}}_k),\bm{\theta})) p^{\mathsf{c}}_{\bm{\mathrm{W}}_k}(\bm{\mathrm{w}}_k|\xi_k) \mathsf{d}\bm{\mathrm{w}}_k} \\
    =&\mathsf{Pr}_{\bm{\mathrm{W}}_k\sim p^{\mathsf{c}}_{\bm{\mathrm{W}}_k}(\cdot|\xi_k)}\left\{\bm{\mathrm{W}}_k\in\overline{\mathcal{W}}_{\bm{\mathrm{x}}_k}^{\bm{\theta}}\right\},\\
    \widehat{\mathbb{P}}^+_{k}(\bm{\theta},\xi_k,\mathcal{D}_N):=&\int_{\mathcal{W}} \Scale[0.96]{\mathbb{I}_{1}(q(f(\bm{\mathrm{x}}_k,\bm{\mathrm{w}}_k),\bm{\theta})) \hat{p}^{\mathsf{c}}_{\bm{\mathrm{W}}_k}(\bm{\mathrm{w}}_k|\xi_k) \mathsf{d}\bm{\mathrm{w}}_k} \\
    =&\mathsf{Pr}_{\bm{\mathrm{W}}_k\sim \hat{p}^{\mathsf{c}}_{\bm{\mathrm{W}}_k}(\cdot|\xi_k)}\left\{\bm{\mathrm{W}}_k\in\overline{\mathcal{W}}_{\bm{\mathrm{x}}_k}^{\bm{\theta}}\right\}.
\end{align*}

Here, the set of $\xi_k$ can share the notation $\mathscr{P}_0:=\mathcal{X}\times\mathscr{V}.$ Replacing $\bm{\mathrm{x}}_0$ and $\rho_0$ in \eqref{eq:convergence_tild_bbP_k_1_p2} by $\bm{\mathrm{x}}_k$ and $\xi_k$, we have that, as $N\rightarrow\infty$, with probability 1,
\begin{equation}
        \label{eq:convergence_tild_bbP_k_k_+_p2}
        \sup_{\bm{\theta}\in\Theta_{\mathsf{c}},\xi_k\in\mathscr{P}_{0}^{\mathsf{c}}}\left|\widehat{\mathbb{P}}^+_{k}(\bm{\theta},\xi_k,\mathcal{D}_N)-\mathbb{P}^+_{k}(\bm{\theta},\xi_k)\right|\rightarrow 0.
\end{equation}

Assume that we have that, as $N\rightarrow\infty$, with probability~1,
\begin{equation}
        \label{eq:convergence_tild_bbP_k_k_assume_p2}
        \sup_{\bm{\theta}\in\Theta_{\mathsf{c}},\rho_{k-1}\in\mathscr{P}_{k-1}^{\mathsf{c}}}\left|\widehat{\mathbb{P}}_k(\bm{\theta},\rho_{k-1},\mathcal{D}_N)-\mathbb{P}_k(\bm{\theta},\rho_{k-1})\right|\rightarrow 0.
\end{equation}

Note that \eqref{eq:convergence_tild_bbP_k_k_assume_p2} has been proved for $k=1$ as shown in \eqref{eq:convergence_tild_bbP_k_1_p2}. We intend to prove that \eqref{eq:convergence_tild_bbP_k_k_assume_p2} holds for $k=k+1$ if \eqref{eq:convergence_tild_bbP_k_k_assume_p2} holds for $k$. Then, the statement (b) can be proven by using the induction method. 

Define $\widehat{\mathbb{P}}_{k+1}(\bm{\theta},\rho_{k},\mathcal{D}_N)$ and $\mathbb{P}_{k+1}(\bm{\theta},\rho_{k})$ by
\begin{align*}
    \widehat{\mathbb{P}}_{k+1}(\bm{\theta},\rho_{k},\mathcal{D}_N)&:=\int_{\mathcal{X}} \Scale[0.95]{\mathbb{I}_{1}(q(\bm{\mathrm{x}}_{k+1},\bm{\theta})) \hat{p}^{\mathsf{c}}_{\bm{\mathrm{X}}_{k+1}}(\bm{\mathrm{x}}_{k+1}|\rho_k) \mathsf{d}\bm{\mathrm{x}}_{k+1}}, \\
    \mathbb{P}_{k+1}(\bm{\theta},\rho_{k})&:=\int_{\mathcal{X}} \Scale[0.95]{\mathbb{I}_{1}(q(\bm{\mathrm{x}}_{k+1},\bm{\theta})) p^{\mathsf{c}}_{\bm{\mathrm{X}}_{k+1}}(\bm{\mathrm{x}}_{k+1}|\rho_k) \mathsf{d}\bm{\mathrm{x}}_{k+1}}.
\end{align*}

They can be computed from the values in the previous step in the following way:
\begin{align*}
    \widehat{\mathbb{P}}_{k+1}(\bm{\theta},\rho_k,\mathcal{D}_N)&=\int_{\mathcal{X}}\int_{\mathcal{W}}\mathbb{I}_{1}(q(\bm{\mathrm{x}}_{k+1},\bm{\theta}))\hat{p}^{\mathsf{c}}_{\bm{\mathrm{W}}_{k}}(\bm{\mathrm{w}}_{k}|\xi_k) \mathsf{d}\bm{\mathrm{w}}_{k} \nonumber \\
    &\ \ \ \ \hat{p}^{\mathsf{c}}_{\bm{\mathrm{X}}_{k}} (\bm{\mathrm{x}}_{k}|\rho_k) \mathsf{d}\bm{\mathrm{x}}_{k} \nonumber \\
    &=\int_{\mathcal{X}} \hat{\mathbb{P}}^+_{k}(\bm{\theta},\xi_k,\mathcal{D}_N) \hat{p}^{\mathsf{c}}_{\bm{\mathrm{X}}_{k}}(\bm{\mathrm{x}}_{k}|\rho_k)\mathsf{d}\bm{\mathrm{x}}_{k}, \\
    \mathbb{P}_{k+1}(\bm{\theta},\rho_k)&=\int_{\mathcal{X}}\int_{\mathcal{W}}\mathbb{I}_{1}(q(\bm{\mathrm{x}}_{k+1},\bm{\theta}))p^{\mathsf{c}}_{\bm{\mathrm{W}}_{k}}(\bm{\mathrm{w}}_{k}|\xi_k) \mathsf{d}\bm{\mathrm{w}}_{k} \nonumber \\
    &\ \ \ \ p^{\mathsf{c}}_{\bm{\mathrm{X}}_{k}}(\bm{\mathrm{x}}_{k}|\rho_k) \mathsf{d}\bm{\mathrm{x}}_{k} \nonumber \\
    &=\int_{\mathcal{X}} \mathbb{P}^+_{k}(\bm{\theta},\bm{\mathrm{x}}_k) p^{\mathsf{c}}_{\bm{\mathrm{x}}_{k}}(\xi_{k}|\rho_k)\mathsf{d}\bm{\mathrm{x}}_{k}. \nonumber
\end{align*}

By \eqref{eq:convergence_tild_bbP_k_k_+_p2} and \eqref{eq:convergence_tild_bbP_k_k_assume_p2}, for any given $\varepsilon,$ there exists $\overline{N}_{\varepsilon}$ such that, if $N>\overline{N}_{\varepsilon}$, with probability 1, we have
\begin{equation}
        \label{eq:convergence_tild_bbP_k_k_+_p2_varepsilon}
        \sup_{\bm{\theta}\in\Theta_{\mathsf{c}},\xi_k\in\mathscr{P}_{0}^{\mathsf{c}}}\left|\widehat{\mathbb{P}}^+_{k}(\bm{\theta},\xi_k,\mathcal{D}_N)-\mathbb{P}^+_{k}(\bm{\theta},\xi_k)\right|<\frac{\varepsilon}{2}.
\end{equation}

Write the compact set $\mathscr{P}_{0}^{\mathsf{c}}$ by $\mathscr{P}_{0}^{\mathsf{c}}=\mathcal{X}_{\mathsf{c}}\times\mathscr{V}_{\mathsf{c}}$, where $\mathcal{X}_{\mathsf{c}}\subset\mathcal{X},\ \mathscr{V}_{\mathsf{c}}\subset\mathscr{V}.$ 
Since there exists a linear transformation (continuous mapping) from $\mathscr{P}_{0}^{\mathsf{c}}$ to $\mathcal{X}_{\mathsf{c}}$, $\mathcal{X}_{\mathsf{c}}$ is also compact.

Let $\bm{\mathrm{E}}_k(\bm{\theta},\xi_k,\mathcal{D}_N):=\widehat{\mathbb{P}}^+_{k}(\bm{\theta},\xi_k,\mathcal{D}_N)-\mathbb{P}^+_{k}(\bm{\theta},\xi_k)$ and $\bm{\mathrm{R}}(\bm{\theta},\rho_k,\mathcal{D}_N):=\widehat{\mathbb{P}}_{k+1}(\bm{\theta},\rho_k,\mathcal{D}_N) - \mathbb{P}_{k+1}(\bm{\theta},\rho_k).$ 
By the iteration computation of $\widehat{\mathbb{P}}_{k+1}(\bm{\theta},\rho_k,\mathcal{D}_N)$ and $\mathbb{P}_{k+1}(\bm{\theta},\rho_k)$, we have
\begin{align*}
    \bm{\mathrm{R}}(\bm{\theta},\rho_k,\mathcal{D}_N)&=\int_{\mathcal{X}} \hat{\mathbb{P}}^+_{k}(\bm{\theta},\xi_k,\mathcal{D}_N) \hat{p}^{\mathsf{c}}_{\bm{\mathrm{X}}_{k}}(\bm{\mathrm{x}}_{k}|\rho_k) \mathsf{d}\bm{\mathrm{x}}_{k} -\\
    &\ \ \int_{\mathcal{X}} \mathbb{P}^+_{k}(\bm{\theta},\xi_k) p^{\mathsf{c}}_{\bm{\mathrm{X}}_{k}}(\bm{\mathrm{x}}_{k}|\rho_k)\mathsf{d}\bm{\mathrm{x}}_{k} \\
    &=\int_{\mathcal{X}}\mathbb{P}^+_{k}(\bm{\theta},\xi_k)\hat{p}^{\mathsf{c}}_{\bm{\mathrm{X}}_{k}}(\bm{\mathrm{x}}_{k}|\rho_k) \mathsf{d}\bm{\mathrm{x}}_{k}-\\
    &\ \ \int_{\mathcal{X}} \mathbb{P}^+_{k}(\bm{\theta},\xi_k) p^{\mathsf{c}}_{\bm{\mathrm{X}}_{k}}(\bm{\mathrm{x}}_{k}|\rho_k)\mathsf{d}\bm{\mathrm{x}}_{k}+ \\
    &\ \ \int_{\mathcal{X}}\bm{\mathrm{E}}_{k}(\bm{\theta},\xi_k)\hat{p}^{\mathsf{c}}_{\bm{\mathrm{X}}_{k}}(\bm{\mathrm{x}}_{k}|\rho_k) \mathsf{d}\bm{\mathrm{x}}_{k} \\
    &=\int_{\mathcal{X}_{\mathsf{c}}}\mathbb{P}^+_{k}(\bm{\theta},\xi_k)\hat{p}^{\mathsf{c}}_{\bm{\mathrm{X}}_{k}}(\bm{\mathrm{x}}_{k}|\rho_k) \mathsf{d}\bm{\mathrm{x}}_{k}-\\
    &\ \ \int_{\mathcal{X}_{\mathsf{c}}} \mathbb{P}^+_{k}(\bm{\theta},\xi_k) p^{\mathsf{c}}_{\bm{\mathrm{X}}_{k}}(\bm{\mathrm{x}}_{k}|\rho_k)\mathsf{d}\bm{\mathrm{x}}_{k}+ \\
    &\ \ \int_{\mathcal{X}_{\mathsf{c}}}\bm{\mathrm{E}}_{k}(\bm{\theta},\xi_k,\mathcal{D}_N)\hat{p}^{\mathsf{c}}_{\bm{\mathrm{X}}_{k}}(\bm{\mathrm{x}}_{k}|\rho_k) \mathsf{d}\bm{\mathrm{x}}_{k}+ \\
    &\ \ \bm{\mathrm{s}}\left(\mathcal{X}\setminus\mathcal{X}_{\mathsf{c}}\right), \\
\end{align*}
where 
\begin{align*} \bm{\mathrm{s}}\left(\mathcal{X}\setminus\mathcal{X}_{\mathsf{c}}\right):=&\int_{\mathcal{X}\setminus\mathcal{X}_{\mathsf{c}}} \hat{\mathbb{P}}^+_{k}(\bm{\theta},\xi_k,\mathcal{D}_N) \hat{p}^{\mathsf{c}}_{\bm{\mathrm{X}}_{k}}(\bm{\mathrm{x}}_{k}|\rho_k) \mathsf{d}\bm{\mathrm{x}}_{k} -\\
    & \int_{\mathcal{X}\setminus\mathcal{X}_{\mathsf{c}}} \mathbb{P}^+_{k}(\bm{\theta},\xi_k) p^{\mathsf{c}}_{\bm{\mathrm{X}}_{k}}(\bm{\mathrm{x}}_{k}|\rho_k)\mathsf{d}\bm{\mathrm{x}}_{k}.
\end{align*}

Then, if $N>\overline{N}_{\mathsf{r}},$ we have 
\begin{align*}
    \left|\bm{\mathrm{R}}(\bm{\theta},\rho_k,\mathcal{D}_N)\right|&\leq\left|\int_{\mathcal{X}_{\mathsf{c}}}\mathbb{P}^+_{k}(\bm{\theta},\xi_k)\hat{p}^{\mathsf{c}}_{\bm{\mathrm{X}}_{k}}(\bm{\mathrm{x}}_{k}|\rho_k) \mathsf{d}\bm{\mathrm{x}}_{k}-\right.\\
    &\ \ \left.\int_{\mathcal{X}_{\mathsf{c}}} \mathbb{P}^+_{k}(\bm{\theta},\xi_k) p^{\mathsf{c}}_{\bm{\mathrm{X}}_{k}}(\bm{\mathrm{x}}_{k}|\rho_k)\mathsf{d}\bm{\mathrm{x}}_{k}\right|+ \\
    &\ \ \left|\int_{\mathcal{X}_{\mathsf{c}}}\bm{\mathrm{E}}_{k}(\bm{\theta},\xi_k,\mathcal{D}_N)\hat{p}^{\mathsf{c}}_{\bm{\mathrm{X}}_{k}}(\bm{\mathrm{x}}_{k}|\rho_k) \mathsf{d}\bm{\mathrm{x}}_{k}\right|+ \\
    &\ \ \left|\bm{\mathrm{s}}\left(\mathcal{X}\setminus\mathcal{X}_{\mathsf{c}}\right)\right| \allowdisplaybreaks \\
    &<\left|\int_{\mathcal{X}_{\mathsf{c}}}\left(\hat{p}^{\mathsf{c}}_{\bm{\mathrm{X}}_{k}}(\bm{\mathrm{x}}_{k}|\rho_k)-p^{\mathsf{c}}_{\bm{\mathrm{X}}_{k}}(\bm{\mathrm{x}}_{k}|\rho_k)\right)\mathsf{d}\bm{\mathrm{x}}_{k}\right|+ \\
    &\ \ \frac{\varepsilon}{2}+\left|\bm{\mathrm{s}}\left(\mathcal{X}\setminus\mathcal{X}_{\mathsf{c}}\right)\right|
\end{align*}
Note that $|\bm{\mathrm{s}}\left(\mathcal{X}\setminus\mathcal{X}_{\mathsf{c}}\right)|\rightarrow 0$ as $\mu_{\rho_0}\left(\mathscr{P}_0\setminus\mathscr{P}_0^{\mathsf{c}}\right)\rightarrow 0$ (for arbitrary measure $\mu_{\rho_0}$), including all considered probability density are continuous and positive all over $\mathcal{X}.$ 
Besides, by Lemma \ref{lemma:convergence_LS_CDE}, it holds that
$$\left|\int_{\mathcal{X}_{\mathsf{c}}}\left(\hat{p}^{\mathsf{c}}_{\bm{\mathrm{X}}_{k}}(\bm{\mathrm{x}}_{k}|\bm{\mathrm{x}}_0)-p^{\mathsf{c}}_{\bm{\mathrm{X}}_{k}}(\bm{\mathrm{x}}_{k}|\bm{\mathrm{x}}_0)\right)\mathsf{d}\bm{\mathrm{x}}_{k}\right|\rightarrow 0.$$

Thus, we can find a set $\mathcal{X}_{\mathsf{c}}$ that is sufficient large to make 
\begin{equation*}
    |\bm{\mathrm{s}}\left(\mathcal{X}\setminus\mathcal{X}_{\mathsf{c}}\right)|+|\bm{\mathrm{s}}\left(\mathcal{X}\setminus\mathcal{X}_{\mathsf{c}}\right)|<\frac{\varepsilon}{2}.
\end{equation*}
Then, with probability 1, we have
\begin{equation}
\label{eq:R_convergence_pointwise}
    \left|\bm{\mathrm{R}}(\bm{\theta},\rho_k,\mathcal{D}_N)\right|<\varepsilon,
\end{equation}
for all $\rho_k$ and $\bm{\theta}.$ The pointwise convergence of $\widehat{\mathbb{P}}_{k+1}(\bm{\theta},\rho_k,\mathcal{D}_N)$ to $\mathbb{P}_{k+1}(\bm{\theta},\rho_k)$ is proved. Then, by repeating the process from \eqref{eq:convergence_tild_bbP_k_1_p2_pointwise} to \eqref{eq:convergence_tild_bbP_k_1_p2}, we obtain \eqref{eq:convergence_tild_bbP_k_k_assume_p2} for $k=k+1$, which completes the proof of (b).

Note that \eqref{eq:convergence_tild_bbP_k_k} can be obtained directly by directly using \eqref{eq:convergence_tild_bbP_k_k_p1} and \eqref{eq:convergence_tild_bbP_k_k_assume_p2}.

\section{Proof of Theorem \ref{theo:almost_uniform_convergence}}
\label{proof:theo_almost_uniform_convergence}

In this proof, $\bm{\theta}_k$ is shortly written as $\bm{\theta}$ for the simplicity of the notation.
Recall \eqref{eq:R_convergence_pointwise} holds for all given $\rho_{k-1}$, $\bm{\theta}$. 
$\widehat{\mathbb{P}}_{k}(\bm{\theta},\rho_{k-1},\mathcal{D}_N)$ and $\mathbb{P}_{k}(\bm{\theta},\rho_{k-1})$ are both continuous function of $\bm{\theta}$, and $\Theta$ is a compact set. For every $\rho_{k-1}\in\mathscr{P}_{k-1},$ as $N,N_{\mathsf{r}}\rightarrow\infty,$ with probability 1, we have
\begin{equation*}
    % \label{eq:convergence_bbP_x_0_uniform}
    \sup_{\bm{\theta}\in\Theta}\left|\widehat{\mathbb{P}}_{k}(\bm{\theta},\rho_{k-1},\mathcal{D}_N)-\mathbb{P}_{k}(\bm{\theta},\rho_{k-1})\right|\rightarrow 0.
\end{equation*}
Then, by Theorem 3.5 of \cite{Pena}, for every $\rho_{k-1}\in\mathscr{P}_{k-1},$ with probability 1, we have that
\begin{align*}
    & \widetilde{J}_{k,\alpha}(\rho_{k-1},\mathcal{D}_N,\widetilde{\mathcal{X}}_k^{N_\mathsf{r}})\rightarrow J^{*}_{k,\alpha}(\bm{\mathrm{x}}_0),\\
    & \mathbb{D}\left(\widetilde{U}_{k,\alpha}(\rho_{k-1},\mathcal{D}_N,\widetilde{\mathcal{X}}_k^{N_\mathsf{r}}),U_{k,\alpha}(\rho_{k-1})\right)\rightarrow 0,
\end{align*}
which are pointwise convergence on $\mathscr{P}_{k-1}$. By Egorov's theorem \cite[page 57]{Wheeden}, we directly conclude Theorem \ref{theo:almost_uniform_convergence}.

\section{Proof of Lemma \ref{theo:uniform_convergence_M_compact}}
\label{proof:theo_uniform_convergence_M_compact}

Note that $\mathsf{d}\widehat{\mathbb{P}}_k(\bm{\theta}, \rho_{k-1}, \mathcal{D}_N)$ is a random variable because it depends on the randomly extracted data set $\mathcal{D}_N$. 
By Theorem \ref{theo:almost_uniform_convergence}, we know that $\mathsf{d}\widehat{\mathbb{P}}_k(\bm{\theta}, \rho_{k-1}, \mathcal{D}_N)$ uniformly converges to zero on a compact set $\Theta_{\mathsf{c}}\times \mathscr{P}^{\mathsf{c}}_{k-1}$ and thus $\overline{M}$ conveges to zero with probability one. 
For simplicity of notation, we denote $-\overline{M}$ as $X_N$.
Here, we assume that $X_N>0$ leading to $\mathsf{Pr}\left\{X_N>\kappa\right\}=\mathsf{Pr}\left\{X_N>\kappa|X_N>0\right\}.$ 
In the case where $X_N\leq 0$ may hold, it will be $\mathsf{Pr}\left\{X_N>\kappa\right\}\leq\mathsf{Pr}\left\{X_N>\kappa|X_N>0\right\}.$ 
Therefore, the upper bound obtained under the assumption that $X_N > 0$ still holds for the general case where $X_N \leq 0$. 
For simplicity in presenting our proof, we assume $X_N > 0$. 
For any given $\eta>0$, we have 
\begin{align}
    \mathsf{Pr}\left\{X_N>\kappa\right\}&=\mathsf{Pr}\left\{\exp(\eta X_N)>\exp(\eta\kappa)\right\} \label{eq:chernov_Markov_dP_1}  \\
    &\leq\frac{\mathbb{E}\left\{\exp(\eta X_N)\right\}}{\exp(\eta\kappa)}. \label{eq:chernov_Markov_dP_2}
\end{align}

Note that \eqref{eq:chernov_Markov_dP_2} can be obtained from the right side of \eqref{eq:chernov_Markov_dP_1} by directly applying Markov's inequality (\cite[p. 265]{Bertsekas_Probability}) to the exponential. 
The Taylor expansion of the Moment Generating Function (MGF) is written by 
\begin{equation}
    \label{eq:MGF_Talor_Expansion}
    \mathbb{E}\left\{\exp(\eta X_N)\right\}=1+\eta \mathbb{E}\left\{ X_N \right\} + \frac{\eta^2}{2} \mathbb{E}\left\{ X_N^2 \right\}+r_2(\eta),
\end{equation}
where $r_2(\eta)$ accounts for higher-order terms written by
\begin{equation}
    r_2(\eta):=\sum_{i=3}^{\infty} \frac{\eta^i}{i!}\mathbb{E}\{X_N^i\}.
\end{equation}

By Lemma \ref{lemma:convergence_LS_CDE}, from the convergence rat, the asymptotic behavior of $X_N$ can be summarized as
\begin{align}
    \mathbb{E}\left\{ X_N \right\}\sim C_1 N^{\frac{2+\gamma}{4}}, \label{eq:asymptotic_behavior_X_N_1} \\
    \mathbb{E}\left\{ X_N^2 \right\}\sim C_2 N^{\frac{2+\gamma}{2}},\label{eq:asymptotic_behavior_X_N_2}
\end{align}
where $C_1$ and $C_2$ are positive constants that depends on the moments of $X_N$, which in turn makes it dependent on $\bm{\theta}$ and $\rho_{k-1}$.

By substituting \eqref{eq:asymptotic_behavior_X_N_1} and \eqref{eq:asymptotic_behavior_X_N_2} into \eqref{eq:MGF_Talor_Expansion} and then substituting \eqref{eq:MGF_Talor_Expansion} into Markov inequility \eqref{eq:chernov_Markov_dP_2}, we can obtain 
\begin{equation}
    \label{eq:Markov_inequility_after_substitute_1}
    \mathsf{Pr}\left\{X_N>\kappa\right\}\leq \frac{1+\eta C_1 N^{\frac{2+\gamma}{4}}+\frac{\eta^2}{2}C_2 N^{\frac{2+\gamma}{2}}+r_2(\eta)}{\exp(\eta\kappa)}.
\end{equation}

Let $Z:=\eta C_1 N^{\frac{2+\gamma}{4}}+\frac{\eta^2}{2}C_2 N^{\frac{2+\gamma}{2}}$ and apply the inequality $1+Z+r_2(\eta)\leq \exp(Z+r_2(\eta))$ since $Z+r_2(\eta)>0$ and $Z+r_2(\eta)\approx 0$ as $\eta\rightarrow 0,\ N\rightarrow\infty$. 
Therefore, the inequality \eqref{eq:Markov_inequility_after_substitute_1} becomes
\begin{align}
    \mathsf{Pr}\left\{X_N>\kappa\right\}&\leq \exp\left(Z+r_2(\eta)-\eta\kappa\right) \nonumber \\
    &\leq C_3\exp\left(Z-\eta\kappa\right), \label{eq:Markov_inequility_after_substitute_2}
\end{align}
where $C_3 = \sup_{\eta \in (0, \bar{\eta}]} r_2(\eta)$, and $\bar{\eta} < 1$ is chosen to ensure that $C_3$ remains sufficiently small. 
Since \eqref{eq:Markov_inequility_after_substitute_2} holds for all $\eta > 0$, we aim to minimize $Z - \eta \kappa$ to reduce the conservatism of the bound. 
The gradient of $Z - \eta \kappa$ must be zero to achieve optimality. 
Therefore, the optimal value of $\eta$, denoted as $\eta^*$, can be obtained by 
\begin{equation}
\label{eq:optimal_eta_approximation}
    \eta^*=\frac{\kappa-C_1 N^{(2+\gamma)/4}}{C_2 N^{(2+\gamma)/2}}\approx\frac{\kappa}{C_2 N^{(2+\gamma)/2}}.
\end{equation}
The approximation holds when $N$ is sufficiently large. 
Additionally, using this approximation does not violate the inequality. 
It merely increases the bound's conservatism. 
Substituting the approximation \eqref{eq:optimal_eta_approximation} into \eqref{eq:Markov_inequility_after_substitute_2}, 
\begin{align}
    \mathsf{Pr}\left\{X_N>\kappa\right\}&\leq C_3\exp\left(-\frac{\kappa^2}{2C_2 N^{(2+\gamma)/4}}\right) \nonumber \\
    &\leq  C_3\exp\left(-\frac{\kappa^2}{C_4 N^{(2+\gamma)/4}}\right).
    \label{eq:Markov_inequility_after_substitute_3}
\end{align}
Here, $C_3, C_4>0$ are constants depending on the moments of $X_N$. 
Replacing $C_3, C_4$ by $\overline{A}_1,\overline{A}_2$ leads to \eqref{eq:uniform_convergence_M}. 

\section{Proof of Theorem \ref{theo:finite_samples_feasibility}}
\label{proof:theo_finite_samples_feasibility}

In this proof, $\bm{\theta}_k$ is shortly written as $\bm{\theta}$ for the simplicity of the notation. 
We briefly summarize Hoeffding's inequality \cite{Hoeffding}, which is applied to prove Theorem \ref{theo:finite_samples_feasibility}. Let $Y_1,...,Y_N$ be independent random variables, with $\mathsf{Pr}\{Y_j\in[y_{j}^{\mathsf{min}},y_{j}^{\mathsf{max}}]\}=1$, where $y_{j}^{\mathsf{min}}\leq y_{j}^{\mathsf{max}}$ for $j\in [N]$. Then, let $e_i^Y:=Y_i-\mathbb{E}[Y_i]$ and $b_i^Y:=y_{j}^{\mathsf{max}}-y_{j}^{\mathsf{min}}$, if $r>0$, the following holds
\begin{equation*}
    \mathsf{Pr}\Big\{\sum_{i=1}^N e_i^Y\geq rN\Big\}\leq\exp\Big\{-\frac{2N^2r^2}{\sum_{i=1}^N(b_i^Y)^2}\Big\}.
\end{equation*}

According to the definition of $M_{\kappa}(\bm{\theta},\rho_{k-1})$ by \eqref{eq:def_barM}, it includes two parts:
\begin{itemize}
    \item The first part $\mathbb{P}_k(\bm{\theta},\rho_{k-1})-\widehat{\mathbb{P}}_k(\bm{\theta},\rho_{k-1},\mathcal{D}_N)$ depends on the data set $\mathcal{D}_N$ and the values of $\bm{\theta},\ \rho_{k-1}$;
    \item The second part $\kappa=\alpha-\alpha_{\mathsf{s}}$ is deterministic.
\end{itemize}

Let $Y_i=\mathbb{I}_{1}(q(\tilde{\bm{\mathrm{x}}}^{(i)}_{k},\bm{\theta}^{\mathsf{v}}_k(\rho_{k-1})))$ for $i=1,...,N_{\mathsf{r}}$, then $\mathsf{Pr}\{Y_i\in[0,1]\}=1$ and $\mathbb{E}\{Y_i\}=\widehat{\mathbb{P}}_k(\bm{\theta}^{\mathsf{v}}_k(\rho_{k-1}),\rho_{k-1},\mathcal{D}_N)$. When $\bm{\theta}^{\mathsf{v}}_k(\rho_{k-1})$ is infeasible for problem \ref{eq:prob_orig}, it holds that 
\begin{equation}
    \label{eq:bbP_k_theta_v}
    \mathbb{P}_k(\bm{\theta}^{\mathsf{v}}_k(\rho_{k-1}),\rho_{k-1})<1-\alpha.
\end{equation}
If $\bm{\theta}^{\mathsf{v}}_k(\rho_{k-1})$ is feasible for \eqref{eq:prob_approximate}, we have
\begin{equation*}
    \widetilde{\mathbb{P}}_k(\bm{\theta}^{\mathsf{v}}_k(\rho_{k-1}),\rho_{k-1},\mathcal{D}_N,\widetilde{\mathcal{X}}_k^{N_\mathsf{r}})\geq 1-\alpha_{\mathsf{s}}.
\end{equation*}
\begin{equation}
\label{eq:bbI_alpha_s}
    \frac{1}{N_{\mathsf{r}}}\sum_{i=1}^{N_{\mathsf{r}}} \mathbb{I}_{1}(q(\tilde{\bm{\mathrm{x}}}^{(i)}_{k},\bm{\theta}^{\mathsf{v}}_k(\rho_{k-1})))\geq 1-\alpha_{\mathsf{s}}.
\end{equation}
Adding $-\mathbb{P}_k(\bm{\theta}^{\mathsf{v}}_k(\rho_{k-1}),\rho_{k-1})$ and to the left and right sides of \eqref{eq:bbI_alpha_s}, with \eqref{eq:bbP_k_theta_v}, we have
\begin{equation}
\label{eq:bbI_alpha_s_alpha}
    \frac{1}{N_{\mathsf{r}}}\sum_{i=1}^{N_{\mathsf{r}}} \mathbb{I}_{1}(q(\tilde{\bm{\mathrm{x}}}^{(i)}_{k},\bm{\theta}^{\mathsf{v}}_k(\rho_{k-1})))-\mathbb{P}_k(\bm{\theta}^{\mathsf{v}}_k(\rho_{k-1}),\rho_{k-1})\geq \alpha-\alpha_{\mathsf{s}}.
\end{equation}
Adding the term $$\mathbb{P}_k(\bm{\theta}^{\mathsf{v}}_k(\rho_{k-1}),\rho_{k-1})-\widehat{\mathbb{P}}_k(\bm{\theta}^{\mathsf{v}}_k(\rho_{k-1}),\rho_{k-1},\mathcal{D}_N)$$ to both sides of \eqref{eq:bbI_alpha_s_alpha}, using $$M(\bm{\theta}^{\mathsf{v}}_k(\rho_{k-1}),\rho_{k-1}):=\mathbb{P}_k(\bm{\theta},\bm{\mathrm{x}}_0)-\widehat{\mathbb{P}}_k(\bm{\theta},\bm{\mathrm{x}}_0,\mathcal{D}_N)+\kappa,$$ $$Y_i=\mathbb{I}_{1}(q(\tilde{\bm{\mathrm{x}}}^{(i)}_{k},\bm{\theta}^{\mathsf{v}}_k(\rho_{k-1}))),$$ and $\mathbb{E}\{Y_i\}=\widehat{\mathbb{P}}_k(\bm{\theta}^{\mathsf{v}}_k(\rho_{k-1}),\rho_{k-1},\mathcal{D}_N),$ we obtain
\begin{align*}
    % \label{eq:bbI_M}
    \frac{1}{N_{\mathsf{r}}}\sum_{i=1}^{N_{\mathsf{r}}}\left(Y_i-\mathbb{E}\{Y_i\}\right)&\geq M(\bm{\theta}^{\mathsf{v}}_k(\rho_{k-1}),\rho_{k-1}) \\
    \sum_{i=1}^{N_{\mathsf{r}}}\left(Y_i-\mathbb{E}\{Y_i\}\right)&\geq N_{\mathsf{r}}M(\bm{\theta}^{\mathsf{v}}_k(\rho_{k-1}),\rho_{k-1}).
\end{align*}
$\mathsf{Pr}\left\{\bm{\theta}^{\mathsf{v}}_k(\rho_{k-1})\in\widetilde{\Theta}_{k,\alpha_{\mathsf{s}}}(\rho_{k-1},\mathcal{D}_N,\widetilde{\mathcal{X}}^{N_{\mathsf{r}}}_k)\right\}$ can be equivalently regarded as $$\mathsf{Pr}\left\{\widetilde{\mathbb{P}}_k(\bm{\theta}^{\mathsf{v}}_k(\rho_{k-1}),\rho_{k-1},\mathcal{D}_N,\widetilde{\mathcal{X}}_k^{N_\mathsf{r}})\geq 1-\alpha_{\mathsf{s}}\right\},$$ which can be rewritten as $$\mathsf{Pr}\left\{\widetilde{\mathbb{P}}_k(\bm{\theta}^{\mathsf{v}}_k(\rho_{k-1}),\rho_{k-1},\mathcal{D}_N,\widetilde{\mathcal{X}}_k^{N_\mathsf{r}})\geq 1-\alpha_{\mathsf{s}}\right\}$$ is equivalently written by $$\mathsf{Pr}\left\{\sum_{i=1}^{N_{\mathsf{r}}}\left(Y_i-\mathbb{E}\{Y_i\}\right)\geq N_{\mathsf{r}}M(\bm{\theta}^{\mathsf{v}}_k(\rho_{k-1}),\rho_{k-1})\right\}.$$ Due to Hoeffding's inequality, we know that $$\mathsf{Pr}\left\{\sum_{i=1}^{N_{\mathsf{r}}}\left(Y_i-\mathbb{E}\{Y_i\}\right)\geq N_{\mathsf{r}}M(\bm{\theta}^{\mathsf{v}}_k(\rho_{k-1}),\rho_{k-1})\right\}$$ is bounded by $\exp\left\{-2N_{\mathsf{r}}M^2(\bm{\theta}^{\mathsf{v}}_k(\rho_{k-1}),\rho_{k-1})\right\}$ if $M(\bm{\theta}^{\mathsf{v}}_k(\rho_{k-1}),\rho_{k-1})\geq 0$ holds. 
Note that if $\overline{M}$ defined by \eqref{eq:def_barM} is not smaller than $-\kappa,$ then $M(\bm{\theta}^{\mathsf{v}}_k(\rho_{k-1}),\rho_{k-1})\geq 0$ holds. 
The probability of having $\overline{M}<-\kappa$ is bounded by \eqref{eq:uniform_convergence_M}. 
Then, replacing $M(\bm{\theta}^{\mathsf{v}}_k(\rho_{k-1}),\rho_{k-1})$ by $\overline{M}$ in $\exp\left\{-2N_{\mathsf{r}}M^2(\bm{\theta}^{\mathsf{v}}_k(\rho_{k-1}),\rho_{k-1})\right\}$ and using Boole's inequality, we have \eqref{eq:finite_samples_feasibility}.

\end{document}